\documentclass[orivec]{llncs}
\usepackage{etex}
\usepackage[utf8]{inputenc}
\usepackage[T1]{fontenc}
\usepackage{amsmath}     %,stmaryrd,mathpartir}
\usepackage{amssymb}
\usepackage[]{graphicx} 
\usepackage{paralist} 
\usepackage{enumitem}
\usepackage{xspace}
\usepackage{url}
\usepackage[inference]{semantic}
\usepackage{fancyhdr}
\usepackage{epsfig}
\usepackage{latexsym}
\usepackage{lmodern}
\usepackage{microtype}
\usepackage{lastpage} 
\usepackage{hyperref}
\usepackage{soul}
\usepackage{graphicx}
\newcommand{\Rand}[2]{#1 \stackrel{\raisebox{-0.25ex}[-0.25ex]{\tiny $\mathdollar$}}{\raisebox{-.25ex}[.25ex]{$\leftarrow$}} #2}

\newcommand\redout{\bgroup\markoverwith
{\textcolor{red}{\rule[.5ex]{2pt}{0.4pt}}}\ULon}

\usepackage{listings}
\lstdefinestyle{customhaskell}{
    language=Haskell,
    showstringspaces=false,
    basicstyle=\footnotesize\rmfamily,
    keywordstyle=\bfseries\color{green!40!black},
    commentstyle=\itshape\color{blue!40!black},
    identifierstyle=\color{black},
    stringstyle=\color{orange!40!black},
}
\usepackage[dvipsnames*,svgnames]{xcolor}
\hypersetup{colorlinks=true,allcolors=DarkBlue,linkcolor=black}

\usepackage{easyReview}

\usepackage{tikz}
\usetikzlibrary{decorations.text, positioning}
\usetikzlibrary{arrows.meta}

\let\origtheorem\theorem
\let\origproof\proof
\usepackage{z-eves}
\let\theorem\origtheorem
\let\proof\origproof

\newcommand{\eqdef}{\stackrel{{\rm def}}{=}}

\newcommand{\tap}[4]{\mbox{$\mathit{#1}(#2,#3,#4)$}}

\newcommand{\step}[1]{\mathbin{\lower0.55ex\hbox{$\lhook\joinrel\xrightarrow{#1}$}}}

\newlength{\bcextramargin}
\newenvironment{changemargin}[2]{\begin{list}{}{% 
\setlength{\topsep}{0pt}% 
\setlength{\leftmargin}{0pt}% 
\setlength{\rightmargin}{0pt}% 
\setlength{\listparindent}{\parindent}% 
\setlength{\itemindent}{\parindent}% 
\setlength{\parsep}{0pt plus 1pt}% 
\addtolength{\leftmargin}{#1}% 
\addtolength{\rightmargin}{#2}% 
}\item }{\end{list}} 
\newcommand{\actdefsection}[1]{
\begin{changemargin}{-\bcextramargin}{0pt}
\vspace{1ex}
\noindent
\textbf{{#1}}
\end{changemargin}
}

\newenvironment{absolutelynopagebreak}
  {\par\nobreak\vfil\penalty0\vfilneg
   \vtop\bgroup}
  {\par\xdef\tpd{\the\prevdepth}\egroup
   \prevdepth=\tpd}

\newcommand{\setlog}{$\{log\}$\xspace}

\newcommand{\Return}{\mathsf{return}\ }
\newcommand{\game}[1]{\textsf{Game }\mathsf{#1}}
\newcommand{\Call}[3]{#1 \leftarrow #2\mathsf{(}#3\mathsf{)}}

\newcommand{\adv}{\mathcal{A}}

\begin{document}
\pagestyle{headings}
\bibliographystyle{plain}
%
%\title{Towards a formal verification of the MimbleWimble cryptocurrency protocol} 
\title{A Formal Analysis of the MimbleWimble Cryptocurrency Protocol}
\author{Adri\'an Silveira\inst{1}, Gustavo Betarte\inst{1}, Maximiliano Cristi\'a\inst{2} \and Carlos Luna\inst{1}} 
\institute{InCo, Facultad de Ingenier\'ia, Universidad de la Rep\'ublica, Uruguay. 
   \\ \email{\{gustun,cluna,adrians\}@fing.edu.uy} 
\and CIFASIS, Universidad Nacional de Rosario, Argentina.
 \\ \email{cristia@cifasis-conicet.gov.ar}
}

\maketitle

\begin{abstract}                                          
% \textit{MimbleWimble} is a cryptocurrency protocol providing different security and scalability properties than other protocols.
%In this paper we put forward an approach for addressing the formal certification of the correctness of MimbleWimble and its implementation (\textit{Grin}). 
%
MimbleWimble (MW) is a  privacy-oriented cryptocurrency technology which provides security and scalability properties that distinguish it from other protocols of its kind.
We present and discuss those properties and outline the basis of a model-driven verification approach to address the certification of the correctness of the protocol implementations.
In particular, we propose an idealized model that is key in the described verification process, and identify and precisely state sufficient conditions for our model to ensure the verification of relevant security properties of MW. Since MW is built on top of a consensus protocol, we develop a Z specification of one such protocol and present an excerpt of the \setlog prototype generated from the Z specification. This \setlog prototype can be used as an executable model where simulations can be run. This allows us to analyze the behavior of the protocol without having to implement it in a low level programming language.
Finally, we analyze the \texttt{Grin} and \texttt{Beam} implementations of MW in their current state of development. \\
%particular implementation of  the protocol.

\textbf{Keywords:} Cryptocurrency, MimbleWimble, Idealized model, Formal verification, Security. 

\end{abstract}
\section{Introduction}
\label{sec:intro}
Cryptocurrency protocols deal with virtual money so they are a valuable target for highly skilled attackers. Several attacks have already been mounted against cryptocurrency systems, causing irreparable losses of money and credibility (e.g. \cite{dao}). For this reason the cryptocurrency community is seeking approaches, methods, techniques and development practices that can reduce the chances of successful attacks. One such approach is the application of formal methods to software implementation. In particular, the cryptocurrency community is showing interest in formal proofs and formally certified implementations \cite{rosu:OASIcs:2020:13414,10.1145/3437378.3437879} . 

MimbleWimble (MW) is a privacy-oriented cryptocurrency technology encompassing  security and scalability properties that distinguish it from other technologies of its kind. 
MW was first proposed in 2016~\cite{mimbleimble-wp}.
The idea was then further developed by Poelstra \cite{poelstra16}. 
In MW,  unlike Bitcoin~\cite{bitcoin}, there exists no concept of address and all the transactions are confidential. In this paper we outline an approach based on formal software verification aimed at formally verifying the basic mechanisms of MW and its implementations \cite{mimbleimble-doc,mw-beam}.

We put forward a model-driven verification approach where security issues that pertain to the realm of critical mechanisms of the MW protocol are explored on an idealized model of this system. Such model abstracts away the specifics of any particular implementation, and yet provides a realistic setting. Verification is then performed on more concrete models, where low level mechanisms are specified. Finally the low level model is proved to be a correct implementation of the idealized model.

\emph{Security (idealized) models} have played an important role in the design and evaluation of high assurance security systems. Their importance was already pointed out in the Anderson report~\cite{Anderson:1972}. The paradigmatic Bell-LaPadula model~\cite{Bell:LaPadula}, conceived in 1973, constituted the first big effort on providing a formal setting in which to study and reason on confidentiality properties of data in time-sharing mainframe systems. 
%
%\emph{State machines}, in turn, are a powerful tool that can be used for modeling many aspects of computing systems. In particular, they 
\emph{State machines} can be employed as the building block of
a security model. The basic features of a state machine model are the concepts of state and state change. 
A \emph{state} is a representation of the system under study at a given time, which should capture those aspects of the system that are relevant to the analyzed problem. 
State changes are modeled by a state transition function that defines the next state based on the current state and input. 
If one wants to analyze a specific safety property of a system using a state machine model, one must first specify what it means for a state to satisfy the property, and then check if all state transitions preserve  it.
Thus, state machines can be used to model the enforcement of a security policy. % on a system.

%In the rest of the  paper we provide  a very brief description of MW, the building blocks of a formal idealized model of the computational behaviour of MW and a brief account of the verification activities we are putting in place in order to verify the protocol and its implementation. An extended version of this paper is available on arXiv \cite{arXiv-MW}.
%\paragraph{{\bf Related work}}
\subsubsection*{{\bf Related work}}
Developers of cryptocurrency software have already shown interest in 
%should not be scared of 
using mathematics as a tool to describe software. In fact, Nakamoto uses maths in his seminal paper on Bitcoin \cite{bitcoin} and Wood uses it to describe the EVM \cite{wood2014ethereum}. However, these descriptions can not be understood as Formal Methods (FM) because they are not based on standardized notations nor on clear mathematical theories. 

On the other hand the FM community has started to pay attention to cryptocurrency software. Idelberger et al. \cite{DBLP:conf/ruleml/IdelbergerGRS16} proposed to use defeasible logic frameworks such as Formal Contract Logic for the description of smart contracts. Bhargavan et al. \cite{DBLP:conf/ccs/BhargavanDFGGKK16} compile \textsc{Solidity} programs into a verification-oriented functional language where they can verify source code. Luu et al. \cite{DBLP:conf/ccs/LuuCOSH16} use the \textsc{Oyente} tool to find and detect vulnerabilities in smart contracts. Hirai \cite{DBLP:conf/fc/Hirai17} uses \textsc{Lem} to formally specify the EVM; Grishchenko, Maffei and Schneidewind \cite{DBLP:conf/post/GrishchenkoMS18} also formalize the EVM but in \textsf{F*}; and Hildenbrandt et al. do the same but with the reachability logic system known as $\mathbb{K}$. P\^irlea and Sergey \cite{Pirlea:2018:MBC:3176245.3167086} present a \textsc{Coq} \cite{coq-manual,coqart} formalization of a blockchain consensus protocol where some properties are formally verified. 

More recently, Rosu \cite{rosu:OASIcs:2020:13414} presented academic and commercial results in developing blockchain languages and virtual machines that come directly equipped with formal analysis and verification tools. 
Hajdu et al. \cite{hajdu2020formal} developed a source-level approach for the formal specification and verification of \textsf{Solidity} contracts with the primary focus on events. 
Santos Reis et al. \cite{santosreis_et_al:OASIcs:2020:13417} introduced Tezla, an intermediate representation of \textsf{Michelson} smart contracts that eases the design of static smart contract analysers.
In \cite{boyd_et_al:OASIcs:2020:13418}, Boyd et al. presented a blockchain model in \textsf{Tamarin}, that is useful for analyzing certain blockchain based protocols.
On the other hand, Garfatta et al. \cite{10.1145/3437378.3437879} described a general overview of the different axes investigated actually by researchers towards the (formal) verification of \textsf{Solidity} smart contracts. 

Additionally, Metere and Dong \cite{10.1007/978-3-319-65127-9_22} present a mechanised formal verification of the Pedersen commitment protocol using \textsc{EasyCrypt} \cite{DBLP:conf/fosad/BartheDGKSS13} and Fuchsbaue et al. \cite{DBLP:conf/eurocrypt/FuchsbauerOS19} introduce an abstraction for the analysis of some security properties of MW.
 Our work assumes some of these results to formalize and analyze the MW protocol, to then propose a methodology to verify their implementations.
 
%Recently, in \cite{DBLP:journals/corr/abs-1908-00591} Betarte et al. outline some formal methods related techniques whose application the cryptocurrency community should take into consideration. Since integrating formal methods into the development process of highly innovative domains cannot be done all at once, the authors conclude that the cryptocurrency community should \emph{gradually} adopt formal methods. Specifically, they propose to start the adoption process by using formal specification languages based on set theory and first-order logic. 
 
Finally, in Betarte et al.~\cite{DBLP:journals/corr/abs-1908-00591} we outline some formal methods related techniques that we consider particularly useful for cryptocurrency software. %We include relevant case studies in the area of cryptocurrency. 
We present some guidelines for the adoption of formal methods in cryptocurrency software projects. We argue that set-based formal modeling (or specification), simulation, prototyping and automated proof can be applied before considering more powerful approaches such as code formal verification. In particular, we show excerpts of a set-based formal specification of a consensus protocol and of the Ethereum Virtual Machine. We also exhibit that prototypes can be generated from these formal models and simulations can be run on them. By last, we show that test cases can be generated from the same models and how automated proofs can be used to evaluate the correctness of these models. The work we present here closely follows the approach of~\cite{DBLP:journals/corr/abs-1908-00591}.

%\paragraph{{\bf Contribution}}
\subsubsection*{{\bf Contribution}}
%\alert{
%In this paper, we present elements that comprise the essential steps towards the development of an exhaustive formalization of the MW cryptocurrency protocol and the analysis of its properties. 
%We also introduce and discuss the basis of a model-driven verification approach to address the certification of the correctness of a 
%protocol's implementation.
%The proposed idealized model in this paper is key in the verification process and it constitutes the main contribution, together with the analysis of its important properties.
%In particular, we determine sufficient conditions on our model to ensure the verification of relevant security properties of MW.}

%\alert{This article builds upon and extends a previously published paper in AIBlock 2020~\cite{BetarteCLSZ20}. A  preliminary version of that work is also available on arXiv~\cite{DBLP:journals/corr/abs-1907-01688}.
%}

%\highlight{This article builds upon and extends a previously published paper in AIBlock 2020~\cite{BetarteCLSZ20}
%}

This article builds upon and extends a previously published paper in AIBlock 2020~\cite{BetarteCLSZ20}\footnote{A  preliminary version of that work is also available on arXiv~\cite{DBLP:journals/corr/abs-1907-01688}.}. In that paper, we present elements that comprise the essential steps towards the development of an exhaustive formalization of the MW cryptocurrency protocol and the analysis of some of its properties. The proposed idealized model constitutes the main contribution together with the analysis of the  essential properties it is shown to verify. We have also introduced and discussed the basis of a model-driven verification approach to address the certification of the correctness of a protocol's implementation.

In the  present paper we extend both the definition of the MimbleWimble protocol and the idealized model. In particular, the formal definition and discussion of the notion of commitment scheme in Section \ref{sec:commitment_scheme} is completely new. We have also extended the section where we study the security properties of MW, incorporating the discussion on the security properties of Pedersen commitments in Section \ref{secc:sec_prop_pedersen_commitment}. We study the strength of the scheme regarding the main security properties a cryptocurrency protocol must have.
The sections Model-driven verification and Mimblewimble implementations are also new. 
In the first one, since MW is built on top of a consensus protocol, we develop a Z specification of one such protocol and present an excerpt of the \setlog prototype generated from the Z specification. This \setlog prototype can be used as an executable model where simulations can be run. This allows us to analyze the behavior of the protocol without having to implement it in a low level programming language. Finally, we compare two MW implementations, \texttt{Grin}~\cite{mimbleimble-doc} and \texttt{Beam} \cite{mw-beam}, with our model and we discuss some features that set them apart. 
%\hlc{agregue un poco mas, ver si es suficiente}

%\paragraph{{\bf Organization of the paper}}
\subsubsection*{{\bf Organization of the paper}}
The rest of the paper is organized as follows. 
Section~\ref{sec:background} provides a brief description of MW. 
Section \ref{sec:model} describes the building blocks of a formal idealized model (abstract state machine) of the computational behaviour of MimbleWimble. Sections~\ref{sec:verification-MW} and \ref{sec:mdd} provide a account of the verification activities we are putting in place in order to verify the protocol and its implementation.
Then, Section \ref{sec:implementations} analyzes the \texttt{Grin} and \texttt{Beam} implementations of MW in their current state of development.
Final remarks and directions for future work are presented in Section \ref{sec:conclusion}.

\section{The MimbleWimble protocol} \label{sec:background}
\label{MW}
%Transactions are at the core of the MimbleWimble protocol and they constitute a derivation of what are known as confidential transactions \cite{GMaxwell:conftransac,AGibson:conftransac}. 
%A confidential transaction allows a sender to encrypt the amount of bitcoins he wants to send by using blinding factors, which are values chosen by the sender that are  used to encrypt bitcoin amounts in a transaction. In a confidential transaction only the two parties involved know the amount of bitcoins being transacted. However, onlookers can still ensure that the transaction is valid by comparing the number of inputs and outputs; if both are the same, then the transaction will be considered valid. Such a procedure ensures that no bitcoins have been created from nothing and is key in preserving the integrity of the system.
%
%Mimblewimble transactions function in a similar way, except  the recipient of a transaction randomly selects a range of blinding factors provided by the sender. This blinding factor is then used as proof of ownership by the receiver, thus, permitting him to spend the bitcoins.
Confidential transactions \cite{GMaxwell:conftransac,AGibson:conftransac} are at the core of the MW protocol. 
A transaction allows a sender to encrypt the amount of bitcoins by using blinding factors. In a confidential transaction only the two parties involved know the amount of bitcoins being exchanged. However, for anyone observing that transaction it is possible to verify its validity by comparing the number of inputs and outputs; if both are the same, then the transaction will be considered valid. Such procedure ensures that no bitcoins have been created from nothing and is key in preserving the integrity of the system.
In MW transactions, the recipient randomly selects a range of blinding factors provided by the sender, which are then used as proof of ownership by the receiver.

The MW protocol aims at providing the following properties \cite{mimbleimble-wp,mimbleimble-doc}:
\begin{itemize}
%\begin{inparaenum}[i)] 
\item Verification of zero sums without revealing the actual amounts involved in a transaction, which implies confidentiality.
\item Authentication of transaction outputs without signing the transaction.
\item Good scalability, while preserving security, by generating smaller blocks---or better, reducing the size of old blocks, producing a blockchain whose size does not grow in time as much as, for instance, Bitcoin's.
\end{itemize}
%\end{inparaenum} 
%
The first two properties are achieved by relying on Elliptic Curves Cryptography (ECC) operations and properties. %In a sense, 
The third one is a consequence of the first two.

%\paragraph{{\bf Verification of transactions}}
\subsection{Verification of transactions}
%\textbf{Verification of transactions.}
If $v$ is the value of a transaction (either input or output) and $H$ is a point over an elliptic curve, %\footnote{aclarar tecnicamente}, 
then $v.H$ encrypts $v$ because it is assumed to be computationally hard to get $v$ from $v.H$ if we only know $H$. However, if $w$ and $z$ are other values such that $v + w = z$, then if we only have the result of encrypting each of them with $H$ we are still able to verify that equation. Indeed:
$$
v + w = z \iff v.H + w.H = z.H
$$
due to simple properties of scalar multiplication over groups. Therefore, with this simple operations, we can check sums of transactions amounts without knowing the actual amounts. 

Nevertheless, say some time ago we have encrypted $v$ with $H$ and now we see $v.H$, then we know that it is the result of encrypting $v$. In the context of blockchain transactions this is a problem because once a block holding $v.H$ is saved in the chain it will reveal all the transactions of $v$ coins. For such problems, MW encrypts $v$ as $r.G + v.H$ where $r$ is a scalar and $G$ is another point in $H$'s elliptic curve, $r$ is called \emph{blinding factor} and $r.G + v.H$ is called \emph{Pedersen commitment}. By using Pedersen commitments, MW allows the verification of expressions such as $v + w = z$ providing more privacy than the standard scheme. In effect, if $v + w = z$ then we choose $r_v$, $r_w$ and $r_z$ such that $r_v.G + r_w.G = r_z.G$ and so the expression is recorded as:
$$
\overbrace{(r_vG + v.H)}^v + \overbrace{(r_w.G + w.H)}^w = 
\overbrace{r_z.G + z.H}^z
$$
making it possible for everyone to verify the transaction without knowing the true values.

%\paragraph{{\bf Authentication of transactions}}
\subsection{Authentication of transactions}
%\textbf{Authentication of transactions.}
Consider that Alice has received $v$ coins and this was recorded in the blockchain as $r.G + v.H$, where $r$ was chosen by her to keep it private. Now she wants to transfer these $v$ coins to Bob. As a consequence, Alice looses $v$ coins and Bob receives the very same amount, which means that the transaction adds to zero: $r.G + v.H - (r.G + v.H) = 0.G - 0.H$. However, Alice now knows Bob's blinding factor because it must be the same chosen by her (so the transaction is balanced). In order to protect Bob from being stolen by Alice, MW allows Bob to add his blinding factor, $r_B$, in such a way that the transaction is recorded as:
$$
(r + r_B).G + v.H - (r.G + v.H) = r_B.G - 0.H
$$
although now it does not sum zero. However, this \emph{excess value} is used as part of an authentication scheme. Indeed, Bob uses $r_B$ as a private key to sign the empty string ($\epsilon$). This signed document is attached to the transaction so in the blockchain we have:
\begin{itemize}
%\begin{inparaenum}[1)] 
\item Input: $I$.
\item Output: $O$.
\item Bob's signed document: $S$.
\end{itemize}
%\end{inparaenum} 
This way, the transaction is valid if the result of decrypting $S$ with $I - O$ (in the group generated by $G$) yields $\epsilon$. If $I - O$ does not yield something in the form of $r_B.G - 0.H$, then $\epsilon$ will not be recovered and so we know there is an attempt to create money from thin air or there is an attempt to steal Bob's money.
%
%\subsubsection*{Good scalability} 
%We will not delve in this issue since it does not constitute one of the primitive security properties of MimbleWimble.
% \subsection{Main characteristics}
% \subsection{Crypto setting}
% \subsection{Transactions and Chain state}
% \subsection{Expected properties}

\subsection{Commitment Scheme}
\label{sec:commitment_scheme}

A commitment scheme \cite{DBLP:reference/crypt/Crepeau05} is a two-phase cryptographic protocol between two parties: a sender and a receiver. At the end of the commit phase the sender is committed to a specific value that he cannot change later and the receiver should have no information about the committed value. 

A non-interactive commitment scheme \cite{BunzBullet} can be defined as follows:

\begin{definition} [Non-interactive Commitment Scheme]
\label{def:commitment_scheme}
A non-interactive commitment scheme $\zeta (Setup, Com)$ consists of two probabilistic polynomial time algorithms, $Setup$ and $Com$, such that:
\begin{itemize}
%\item $Setup$ generates public parameters for the scheme: $pp \leftarrow Setup(1^{\lambda})$, for the security parameter $\lambda$. 
\item $Setup$ generates public parameters for the scheme depending on the security parameter $\lambda$.
%\item $Com$ is the commitment algorithm: $Com: M_{pp} \times R_{pp} \rightarrow C_{pp}$, where $M$ is the message space, $R$ the randomness space and $C$ the commitment space determined by $pp$.
\item $Com$ is the commitment algorithm: $Com: M \times R \rightarrow C$, where $M$ is the message space, $R$ the randomness space and $C$ the commitment space.
For a message $m \in M$, the algorithm draws uniformly at random $r\leftarrow R$ and computes the commitment $com \leftarrow Com(m,r)$.
\end{itemize}
\end{definition} 

We have simplified the notation, but it is important to keep in mind that all the sets depend on the public parameters, in particular, the commitment algorithm.

It is said that the commitment scheme is homomorphic if:
$$
 \textit{for all } m_1,m_2 \in M, r_1,r_2 \in R \textit{:}
$$
$$
Com(x_1,r_1) + Com(x_2, r_2) = Com(x_1 + x_2, r_1 + r_2)
$$

In other words, $Com$ is additive in both parameters.

Transactions in MW are derived from confidential transactions \cite{GMaxwell:conftransac}, which are enabled by Pedersen commitments with homomorphic properties over elliptic curves.
We define the non-interactive Pedersen commitment scheme we will use in our model, based on Definition \ref{def:commitment_scheme}, as follows: 

\begin{definition} [Pedersen Commitment Scheme with Elliptic Curves]
\label{def:pedersen_commitment}
As in Definition \ref{def:commitment_scheme}, let $M$ and $R$ be the finite field $\mathbb{F}_n$ and let $C$ be the set of points determined by an elliptic curve $\mathcal{C}$ of prime order $n$. 

The probabilistic polynomial time algorithms are defined as:
\begin{itemize}
\item $Setup$ generates the order $n$ (dependent on the security parameter $\lambda$) and two generator points $G$ and $H$ on the elliptic curve $\mathcal{C}$ of prime order $n$ whose discrete logarithms relative to each other are unknown.  
\item $Com(v,r) = r.G + v.H$, with $v$ the transactional value and $r$ the blinding factor choosen randomly in $\mathbb{F}_n$.
\end{itemize}
\end{definition} 

Security properties of this commitment scheme (for MW) will be analyzed in Section \ref{secc:sec_prop_pedersen_commitment}.

%\input{fm}
%\section{Verified implementation}
\section{Idealized model of MimbleWimble-based blockchain}
\label{sec:model}

The basic elements of our model are transactions, blocks and chains. Each node in the blockchain maintains a local state. The main components are the local copy of the chain and the set of transactions waiting to be validated and added to a new block. Moreover, each node keeps track of unspent transaction outputs (UTXOs). Properties such as zero-sum and the absence of double spending in blocks and chains must be proved for local states. 
The blockchain global state can be represented as a mapping from nodes to local states. For global states, we can state and prove properties for the entire system like, for instance,  correctness of the consensus protocol. 

\subsection{Transactions}

Given two fixed generator points $G$ and $H$ on the elliptic curve $\mathcal{C}$ of prime order $n$ (whose discrete logarithms relative to each other are unknown), we define a single transaction between two parties as follows:

\begin{definition} [Transaction] A single transaction $t$ is a tuple of type:
\label{def:transac}
$$
Transaction \eqdef \{i : I^{\star},\ o : O^{\star}, tk : TxKernel, tko : KOffset \}
$$
with $X^{\star}$ representing the lists of elements of type $X$ and where:
\begin{itemize}
\item $i = (c_1, ..., c_n)$ and $o = (o_1, ..., o_m)$ are the lists of inputs and outputs. Each input $c_i$ and output $o_i$ are points over the curve $C$ and they are the result of computing the Pedersen commitment $r.G + v.H$ with $r$ the blinding factor and $v$ the transactional value in the finit field $\mathbb{F}_n$.
%\hl{esto yo no lo entiendo; no entiendo la relacion entre r, G, v y H con los $c_i$ y los $o_i$; mas abajo parece ser que cada $o_i$ es de la forma $r.G + v.H$, en ese caso trataria de ser mas explicito sobre que forma tiene cada $o_i$}
\item $tk = \{rp, ke, \sigma\}$ is the transaction kernel where:
				\begin{itemize}
					\item $rp$ is a list of range proofs of the outputs.
					\item $ke$ is the transaction excess represented by $(\sum_{1}^{m}{r'} - \sum_{1}^{n}{r} - tko).G$.
					\item $\sigma$ is the kernel signature\footnote{For simplicity, fees are left aside.}.
				\end{itemize}
\item $tko \in \mathbb{F}_n$ is the transaction kernel offset.
%\hl{$\mathbb{F}_n$ es el tipo de los conjuntos finitos de enteros de cardinalidad $n$?; sea lo que fuere me parece que habria que aclarar que es}
\end{itemize}	

\end{definition}

The transaction kernel offset will be used in the construction of a block to satisfy security properties.

% Lo comento porque se repite con la definicion de balanceada

% The transaction excess can be seen as: 
% $$ \sum_{(r',v') \in O^{\star}}{r'.G + v'.H} - \sum_{(r,v) \in I^{\star}}{ r.G + v.H}\ $$
% We say that the transaction is balanced iff $\sum_{}{v'} - \sum_{}{v} = 0$. So, in other words, the excess is $(\sum_{}{r'} - \sum_{}{r})G$.
% The kernel signature proves that the transaction is honestly constructed, in particular that the excess only contains the blinding factor (no money is being created).

% The transaction kernel offset will be used in the construction of a block to satisfy security properties.

\begin{definition}[Ownership] Given a transaction $t$, we say $S$ owns the output $o$ 
if $S$ knows the opening $(r,v)$ for the Pedersen commitment $o = r.G + v.H$.
\end{definition}

The strength of this security definition is directly related to the difficulty of solving the logarithm problem. If the elliptic curve discrete logarithm problem in $C$ is hard then given a multiple $Q$ of $G$, it is computationally infeasible to find an integer $r$ such that $Q=r.G$.

\begin{definition} [Balanced Transaction] 
A transaction $t= \{i, o, tk, tko\}$, with transaction kernel $tk = \{rp, ke, \sigma\}$, is balanced if the following holds:
\label{def:balanced_transac}
$$
\sum_{o_j \in o}{o_j} - \sum_{c_j \in i}{c_j} = ke + tko.G
$$
\end{definition}

A balanced transaction guarantees no money is created from thin air and the transaction was honestly constructed.

\begin{property}[Valid Transaction] 
A {\it transaction $t$ is valid} ($valid\_transaction(t)$) if $t$ satisfies: %the following attributes: % attributes no me convence
\label{prop:valid_transac}

\begin{enumerate}[label=\roman*.]
\item The range proofs of all the outputs are valid.
\item The transaction is balanced.
\item The kernel signature $\sigma$ is valid for the excess.
\end{enumerate}
\end{property}

These three properties have a straightforward interpretation in our model. 
Due to limitations of space, we formalize and analyze in this paper only some of the properties mentioned throughout the document.

\subsection{Aggregate Transactions} Transactions can be aggregated into bigger transactions. A single transaction can be seen as the sending of money between two parties. The following definition represents multiple parties:

\begin{definition} [Aggregate Transaction] An aggregate transaction $tx$ is a tuple of type:
\label{def:transac_agg}
$$
TransacAgg \eqdef \{i : I^{\star},\ o : O^{\star}, tk : TxKernel^{\star}, tko : KOffset \}
$$
with $X^{\star}$ representing the lists of $X$ elements detailed in definition \ref{def:transac}.

\end{definition}

Transactions can be merged non-interactively to contruct an agreggate transaction. 

\begin{definition} [Transaction Join] Given a valid transaction $t_0$ and an aggregate transaction $tx$:
\label{def:transac_join}

$$
t_0 = \{i_0, o_0, tk_0, tko_0 \} \textrm{ and } tx = \{ i , o, tk, tko \}
$$

a new aggregate transaction can be constructed as:
$$
tx = \{ i_0 || i, o_0 || o ,  tk_0 || tk , tko_0 + tko \}
$$
Where $||$ is the list concatenation operator and $+$ is the scalar sum.
\end{definition}

This process, called CoinJoin, can be applied recursively to add more transactions into one aggregate transaction. A single transaction (Definition \ref{def:transac}) can be seen as a particular case where the transaction kernel list cointains a single element. Besides, the ownership of the coins is between two parties. 

The validity of an aggregate transaction is guaranteed by the validity of the transactional parties during the construction process.

\begin{lemma} [Invariant: CoinJoin Validity] 
\label{lemma:inv_coinjoin}
Given a valid transaction $t_0$ and a valid aggregate transaction $tx$ as in Definition \ref{def:transac_agg}.
Let $tx'$ be the result of aggregating $t_0$ into $tx$. Then, $tx'$ is valid.
\end{lemma}

In our model, aggregate transactions and blocks (Definition \ref{def:block}) are the same (without considering headers). We are interested in distinguishing them because the unconfirmed transaction pool will contain aggregate transactions.

Notice that, in an aggregate transaction, an adversary could find out which input cancel which output. They could try all possible permutations and verify if they summed to the transaction excess. The property of transaction unlinkability will be proved over blocks, as we will see in Property \ref{prop_unlinkability}.

\subsection{Unconfirmed Transaction Pool}

The unconfirmed transaction pool (mempool) contains the transactions which have not been confirmed in a block yet.

\begin{definition}[Mempool] A mempool $mp$ is a list of type:
\label{def:mempool}

$$
Mempool \eqdef  AggregateTransaction^{\star}
$$

\end{definition}

\subsection{ Blocks and chains}
\label{sec:blocks_and_chains}

Genesis block $Gen$ is a special block since it is the first ever block recorded in the chain. Transactions can be merged into a {\it block}. We can see a block as a big transaction with aggregated inputs, outputs and transaction kernels. 

\begin{definition}[Block] A {\it Block} $b$ is either the {\it genesis block $\mathit{Gen}$}, or a tuple of type:
\label{def:block}

$$
Block \eqdef \{i : I^{\star},\ o : O^{\star}, \ tks : TxKernel^{\star}, ko : KOffset \}
$$
where:
\begin{itemize}
\item $i = (c_1, ..., c_n)$ and $o = (o_1, ..., o_m)$ are the lists of inputs and outputs of the transactions.
%\hl{idem definicion 1}
\item $tks = (tk_1, ..., tk_t)$ is the list of $t$ transaction kernels.  
\item $ko \in \mathbb{F}_n$ is the block kernel offset which covers	all the transactions of the block.	
\end{itemize}	
\end{definition}

We can say a block is balanced if each aggregated transaction is balanced.

\begin{definition}[Balanced Block] Let $b$ be a block of the form $b= \{i, o, tks, ko\}$ with $tks = (tk_1, ..., tk_t)$ the list of transaction kernels and where 
\label{def_balanced_block}
the j-th item in $tks$ is of the form $tk_j = \{rp_j, ke_j, \sigma_j\}$.
We say the block $b$ is balanced if the following holds:
$$
\sum_{o_j \in o}{o_j} - \sum_{c_j \in i}{c_j} = ko.G + \sum_{ke_j \in tks}{ke_j}
$$
\end{definition}

We assume the genesis block $Gen$ is valid. We define the notion of block validity as follows:

\begin{property} [Valid Block] A block $b$ is valid ($valid\_block(b)$) if $b$ is the genesis block $Gen$ or it satisfies:
\label{prop:valid_block} 
\begin{enumerate}[label=\roman*.]
%\item All inputs come from the UTXO set. %se puede sacar, hay predicado que valida lo mismo al agregar b a chain
\item The block is balanced.
\item For every transaction kernel, the range proofs of all the outputs are valid and the kernel signature $\sigma$ is valid for the transaction excess.
\end{enumerate}

\end{property}

Blocks can be constructed by aggregating transactions as follows:

\begin{definition}[Block Aggregation] Given a valid transaction $t_0$ and a valid block $b$ as follows:
\label{def:aggregate_block}
$$
t_0 = \{ i_0 , o_0, tk_0, tko_0 \} \textrm{ and } b = \{i, o, tks, ko\}
$$
a new block can be constructed as:
$$
b' = \{ i_0 || i, o_0 || o ,  tk_0 || tks , tko_0 + ko \}
$$
where $||$ is the list concatenation operator and $+$ is the scalar sum.
\end{definition}

Block agreggation preserves the validity of blocks; i.e. block validity is invariant w.r.t. block agreggation.

\begin{lemma}[Invariant: Block Validity] 
Given a valid transaction $t_0$ and a valid block $b$ as in Definition \ref{def:aggregate_block}.
\label{lema:inv_val_block}
Let $b'$ be the result of aggregating $t_0$ into $b$. Then, $b'$ is valid.
\end{lemma}

%\begin{Proof}
\subsubsection*{{\bf Proof.}} 
Let $t_0$ be the transaction $t_0 = \{ i_0 , o_0, tk_0, tko_0 \}$ with $tk_0=\{rp_0, ke_0, \sigma_0\}$. 
Let $b$ be the block $b= \{i, o, tks, ko\}$, with $tks = (tk_1, ..., tk_t)$, the list of transaction kernels. 

Applying Definition \ref{def:aggregate_block}, we have that the resulting $b'$ is of the form: 

$$
b' = \{ i', o', tks', ko'\} 
$$ 
$$
\textrm{ with } i' = i_0 || i,\ o' = o_0 || o,\ tks' = (tk_0, tk_1, ..., tk_t),\ ko' = tko_0 + ko
$$

According to Definition \ref{def_balanced_block}, we need to prove the following equality holds for the block $b'$:

$$
\sum_{o_j \in o'}{o_j} - \sum_{c_j \in i'}{c_j} = ko'.G + \sum_{ke_j \in tks'}{ke_j}
$$
Each term can be written as follows:
$$
( \sum_{o_j \in o_0}{o_j} + \sum_{o_j \in o}{o_j} )
- 
( \sum_{c_j \in i_0}{c_j} + \sum_{c_j \in i}{c_j} )
= 
(tko_0 + ko) .G 
+ 
ke_0
+
\sum_{ke_j \in tks}{ke_j}
$$
Rearranging the equality and using algebraic properties on elliptic curves, we have:
$$
( \sum_{o_j \in o_0}{o_j} - \sum_{c_j \in i_0}{c_j} )
+ 
(\sum_{o_j \in o}{o_j} - \sum_{c_j \in i}{c_j} )
= 
( ke_0 + tko_0.G )
+
( ko.G + \sum_{ke_j \in tks}{ke_j} )
$$
Now, we apply the hypothesis concerning the validity of $t_0$ and $b$. In particular, applying Definition \ref{def:balanced_transac} for $t_0$ and Definition \ref{def_balanced_block} for $b$, we have the following equalities are true:
$$
\sum_{o_j \in o_0}{o_j} - \sum_{c_j \in i_0}{c_j} = ke_0 + tko_0.G
$$
and
$$
\sum_{o_j \in o}{o_j} - \sum_{c_j \in i}{c_j} = ko.G + \sum_{ke_j \in tks}{ke_j}
$$
That is exactly what we wanted to prove. 
\qed
%\end{proof}

Notice that proof above is analogous to the proof of CoinJoin Validity (Lemma \ref{lemma:inv_coinjoin}).

\begin{definition} [Chain] A {\it chain} is a non-empty list of {\it blocks}:
$$\mathit{Chain} \eqdef \mathit{Block}^{\star}$$
\end{definition}

For a chain $c$ and a valid block $b$, we can define a predicate 
$\mathit{validate(c, b)}$ representing the fact that is correct to add $b$ to $c$. This relation must verify, for example, that all the inputs in $b$ are present as outputs in $c$, in other words, they are unspent transaction outputs (UTXOs).

%\paragraph{{\bf Validating a chain}}
\subsection{Validating a chain}
\label{secc:validating_chains}

%textbf{Validating a chain.}
The model formalizes a notion of valid state that captures several well-formedness conditions. In particular, every block in the blockchain must be valid.
A predicate $\mathit{validChain}$ can be defined for a chain 
$c = (b_{0}, b_{1}, \dots b_{n})$ by checking that:
\begin{itemize}
%\begin{inparaenum}[1)] 
 \item $b_0$ is a valid genesis block
 \item For every $i\in \{1, \dots n\}$, 
 $\mathit{validate}((b_{0}, \dots, b_{i-1}), b_{i})$
\end{itemize}

%%%%%%%%%%%%%%%%%%%%%%%
The axiomatic semantics of the system are modeled by defining a set of transactions, and providing their semantics as state transformers.  
%The behaviour of actions is specified by pre- and postconditions. 
The behaviour of transactions is specified by a precondition $Pre$ and by a postcondition $Post$:
%$$Pre \subseteq  State \times Transaction \hspace{1cm} Post \subseteq  State \times Transaction \times State$$
$$Pre \subseteq  State \times Transaction$$
$$Post \subseteq  State \times Transaction \times State$$

This approach is valid when considering local (nodes) or global (blockchain) states (of type $State$) and transactions (of type $Transaction$). Different sets of transactions, pre and postcondition are defined to cover local or global state transformations. 
At a general level, $State$ is $Chain$.

%\paragraph{{\bf Executions}}
\subsection{Executions}
\label{secc:executions}
There can be attempts to execute a transaction on a state that does not verify the precondition of that transaction. In the presence of such situation the system answers with a corresponding error code (of type $ErrorCode$). 
Executing a transaction $t$ over a valid state $s$ ($valid\_state(s)$)\footnote{When dealing with global states, $valid\_state$ is $validChain$.}
produces a new state $s'$ and a corresponding answer $r$ (denoted $s\step{t/r}s'$),
where the relation between the former state and the new one is given by the postcondition relation $Post$. 
%\hl{The notation $s \step{t/ok} s'$ may be read as \textit{the execution of the transaction $t$ in a valid state $s$ results in a new state $s'$}. Then, we have the following:}
\begin{displaymath}
\begin{array}{c}
\hspace{2.3cm}\inference[]{$$valid\_state(s)$$ \hspace{.2cm} $$Pre(s, t)$$ \hspace{.2cm} $$Post(s, t, s')$$}{$$s\step{t/ok}s'$$} \\ \\
\hspace{2.3cm}\inference[]{$$valid\_state(s)$$ \hspace{.2cm} $$ErrorMsg(s, t, ec)$$}{$$s\step{t/error(ec)}s$$}\\ \\
\end{array}
\end{displaymath}
\noindent Whenever a transaction occurs for which the precondition holds,
the valid state may change in such a way that the transaction
postcondition is established. 
The notation $s \step{t/ok} s'$ may be read as \textit{the execution of the transaction $t$ in a valid state $s$ results in a new state $s'$}.
However, if the precondition is not satisfied, then the valid state $s$ remains unchanged and the system answer is the error message determined by a relation $ErrorMsg$\footnote{Given a state $s$, a transaction $t$ and an error code $ec$, $\tap{ErrorMsg}{s}{t}{ec}$ holds iff $error~ec$ is an acceptable response when the execution of $t$ is requested on state $s$.}. 
Formally, the possible answers of the system are defined by the type: 
$$\textit{Response} \eqdef ok\ |\ error\ (ec : ErrorCode)$$ 
where  $ok$ is the answer resulting from a successful execution of a transaction.

One-step execution with error management preserves valid states.
\begin{lemma} [Validity is invariant]
\label{lemma:valid-state-correct}
\mbox{}\\
$\begin{array}{l}
\forall\ (s\ s': State)(t:Transaction) (r:Response),\\
valid\_transaction(t) \rightarrow s\step{t/r}s' \rightarrow valid\_state(s')
\end{array}$
\end{lemma}
%The properties in this work are obtained from valid states of the system.
The proof follows by case analysis on $s\step{t/r}s' $. When $Pre(s,t)$ does not hold, $s = s'$. From this equality and $valid\_state(s)$
then $valid\_state(s')$. Otherwise, $Pos(s,t,s')$ must hold and we proceed by case analysis on $t$, considering that $t$ is a valid transaction and $s$ is a valid state.\\

System state invariants, such as state validity, are useful to analyze
other relevant properties of the model. In particular, 
the properties in this work are obtained from valid states of the system.
%%%%%%%%%%%%%%%%%%%%%%%

%Valid states are invariant under execution. The properties in this work are obtained from valid states of the system.

%\paragraph{{\bf Action semantics}}
%\subsection{Action semantics}
%\label{sec:model:actsemI}
%The axiomatic semantics of the system is modeled by defining a set of actions,
%and providing their semantics as state transformers.  
%The behaviour of actions is specified by a precondition and by a postcondition.
%
%This approach is valid when considering local (nodes) or global (blockchain) states and actions. Different set of actions, pre and postcondition are defined to cover local or global state transformations.
%\hly{TBC}

\newtheorem{prop}{Property}

\section{Verification of MimbleWimble}
\label{sec:verification-MW}
%We now proceed to discuss some relevant properties that can be verified \hly{on} our model \sty{guarantees}.
%\hlc{igual revisen esa frase, parece que algo no va}
We now detail some relevant properties that can be verified in our model.
In addition to some of the properties mentioned in previous sections, we include in our research other properties such as those formulated in \cite{Pirlea:2018:MBC:3176245.3167086}, and various security properties considered in \cite{DBLP:conf/eurocrypt/GarayKL15,DBLP:conf/crypto/KiayiasRDO17,DBLP:conf/eurocrypt/FuchsbauerOS19}.

%\paragraph{{\bf Protocol Properties}}
%\subsubsection{Protocol Properties.} %\colorbox{cyan}{downgraded to subsection}}
%\label{sec:properties}
% % % % % capaz que comentar desde acá
%\hlc{cambie a subsubsection}
%\noindent
%\paragraph{{\bf Protocol Properties}}
\subsection{Protocol Properties}
\label{sec:protocol_properties}

The property of \textit{no coin inflation} or \textit{zero-sum} guarantees that no new funds are produced from thin air in a valid transaction. The property can be stated as follows. 

%proved using properties of ECC. This property, and its proof, can be extended to blocks because adding zero-sum Pedersen commitments is a zero-sum Pedersen commitment. 

\begin{lemma} [No Coin Inflation] 
Given a valid transaction $t= \{i, o, tk, tko\}$ with transaction kernel $tk = \{rp, ke, \sigma\}$, 
\label{lemma:no_coin_inflation} then the transaction excess only contains the blinding factor and the kernel offset. 

\end{lemma}

\subsubsection*{{\bf Proof.}} 
We know the transaction $t$ is valid, in particular, the transaction is balanced. 
Applying Definition \ref{def:balanced_transac}, we know that:
$$
\sum_{o_j \in o}{o_j} - \sum_{c_j \in i}{c_j} = ke + tko.G
$$
Using Definition \ref{def:transac}, we start to unfold the terms in the equality:
$$
\sum_{1}^{m}{r'.G + v'.H} - \sum_{1}^{n}{r.G + v.H} = (\sum_{1}^{m}{r'} - \sum_{1}^{n}{r} - tko).G + tko.G
$$
Applying algebraic properties on elliptic curves, we have:
$$
\sum_{1}^{m}{v'.H} - \sum_{1}^{n}{v.H} = (\sum_{1}^{m}{r'.G} - \sum_{1}^{n}{r.G}) - (\sum_{1}^{m}{r'.G} - \sum_{1}^{n}{r.G}) - tko.G + tko.G = 0
$$
Therefore, 
$$
 (v'_1 + ... + v'_m).H - (v_1 + ... + v_n) .H = (v'_1 + ... + v'_m - v_1 - ... - v_n) . H = 0.H = 0
$$
It means that all the inputs and outputs add to zero. In other words, they summed to the commitment to the kernel offset plus the commitment to the excess blinding factor. 
\qed

Thus, we have proved no money is created from thin air and the only ones who knew the blinding factors were the transacting parties when they created the transaction. This means the new outputs will be spendable only by them.

An important feature of MW is the \textit{cut-through} process. The purpose of this process is to erase redundant outputs that are used as inputs within the same block. Let $C$ be some coins that appear as an output in the block $b$. If the same coins appear as an input within the block, then $C$ can be removed from the list of inputs and outputs after applying the  cut-through process. In this way, the only remaining data are the block headers, transaction kernels and unspent transaction outputs (UTXOs).
After applying cut-through to a valid block $b$ it is important to ensure that the resulting block $b'$ is still valid. We can say that the validity of a block should be invariant with respect to the cut-through process.

\begin{lemma} [Invariant: Cut-through Block Validity]
\label{lemma:balanced_transac} 
Let $b$ be a block of the form $b= \{i, o, tks, ko\}$ with $i$ and $o$ the list of inputs and outputs, $tks = (tk_1, ..., tk_t)$ the list of transaction kernels and $ko$ the block kernel offset.
Let $b'= \{i', o', tks, ko\}$ be the resulting block after applying the cut-through process to $b$ where:
\begin{itemize}
\item $i' = i \setminus (i \cap o)$
\item $o' = o \setminus (i \cap o)$
\end{itemize}
Hence, if $b$ is a valid block, then $b'$ is valid too. 

\end{lemma}

\subsubsection*{{\bf Proof.}} 
Let $b$ be the block $b= \{i, o, tks, ko\}$, with $tks = (tk_1, ..., tk_t)$ the list of transaction kernels, where 
the j-th item in $tks$ is of the form $tk_j = \{rp_j, ke_j, \sigma_j\}$. 

Let $r$ be $r = i \cap o = \{r_0, r1, ..., r_n\}$ where we assume $r \neq \emptyset$ because otherwise the lemma holds trivially as $b' = b$.

Let $b'$ be the block $b'= \{i', o', tks, ko\}$, with $tks = (tk_1, ..., tk_t)$, the list of transaction kernels, $i' = i \setminus r$ and $o' = o \setminus r$.

We want to prove that $b'$ is valid. In particular, that $b'$ is balanced. 
According to Definition \ref{def_balanced_block}, we need to prove:

$$
\sum_{o_j \in o'}{o_j} - \sum_{c_j \in i'}{c_j} = ko.G + \sum_{ke_j \in tks}{ke_j}
$$

By hypothesis, we know that $b$ is a valid block. Applying Property \ref{prop:valid_block}, we know that $b$ is balanced.
According to Definition \ref{def_balanced_block}, the following equality holds for block $b$:

$$
\sum_{o_j \in o}{o_j} - \sum_{c_j \in i}{c_j} = ko.G + \sum_{ke_j \in tks}{ke_j}
$$

Applying the definition of $r$, we can rewrite the above equality as follows:

$$
(\sum_{o_j \in o \setminus r}{o_j} + \sum_{o_j \in r}{o_j} ) - (\sum_{c_j \in i \setminus r}{c_j} + \sum_{c_j \in r}{c_j})
 = ko.G + \sum_{ke_j \in tks}{ke_j}
$$

Rearranging the equality, we have:

$$
(\sum_{o_j \in o \setminus r}{o_j} - \sum_{c_j \in i \setminus r}{c_j}) + (\sum_{o_j \in r}{o_j} - \sum_{c_j \in r}{c_j})
 = ko.G + \sum_{ke_j \in tks}{ke_j}
$$

Now, we can observe that we are substracting the sum of all the elements belonging to the same set $r$. Thus, the term is equal to zero.

Then, if we remove the term we have:

$$
\sum_{o_j \in o \setminus r}{o_j} - \sum_{c_j \in i \setminus r}{c_j} = ko.G + \sum_{ke_j \in tks}{ke_j}
$$

By hypothesis, we know that $i' = i \setminus r$ and $o' = o \setminus r$; therefore we can rewrite the above equality as:

$$
\sum_{o_j \in o'}{o_j} - \sum_{c_j \in i'}{c_j} = ko.G + \sum_{ke_j \in tks}{ke_j}
$$

That is exactly what we wanted to prove.
\qed
%\end{proof}

%\paragraph{\it Privacy and Security Properties.} %\colorbox{cyan}{downgraded to subsection}}
%\label{sec:sec_props}
%\noindent
%\paragraph{{\bf Privacy and Security Properties}}
\subsection{Privacy and Security Properties}
\label{sec:privacy_security_properties}
%\textbf{Privacy and Security Properties.} 
In blockchain systems the notion of privacy is crucial: sensitive data should not be revealed over the network. In particular,  it is desirable to ensure properties such as  confidentiality, anonymity and unlinkability of transactions. Confidentiality refers to the property of preventing other participants from knowing certain information about the transaction, such as the amounts and addresses of the owners. Anonymity refers to the property of hiding the real identity from the one who is transacting, while unlinkability refers to the inability of linking different transactions of the same user within the blockchain. 

In the case of MW no addresses or public keys are used; there are only encrypted inputs and outputs. 
%Therefore the communication between the sender and receiver to share the proof of ownership of the coins must be done off-chain using a secure channel. 
Privacy concerns rely on confidential transactions, cut-through and CoinJoin. CoinJoin combines inputs and outputs from different transactions into a single unified transaction. 
%Thus, for a third-party it is difficult to determine which party is making a particular transaction. 
It is important to ensure that the resulting transaction satisfies the validity defined in the model.

The security problem of double spending refers to spending a coin more than once. All the nodes keep track of the UTXO set, so before confirming a block to the chain, the node checks that the inputs come from it. If we refer to our model, that validation is performed in the predicate \textit{validate} mentioned in Section \ref{sec:blocks_and_chains}.  

\subsection{Security properties of Pedersen commitments}
\label{secc:sec_prop_pedersen_commitment} 

In MW transactions, input and output amounts are hidden in Pedersen commitments. In Section \ref{sec:commitment_scheme} we have introduced the definition of a commitment schema (Definition \ref{def:commitment_scheme}). 

A commitment scheme is expected to satisfy the following two security properties:

\begin{itemize}
\item \textit{Hiding}: the receiver, who received the commitment, does not learn anything about the original value.
\item \textit{Binding}: after the commit stage, there is at most one value that the sender can successfully open. 
\end{itemize}

In the cryptocurrencies world, these two properties should be understood this way:

\begin{itemize}
\item \textit{Hiding}: a commitment scheme is used to keep the transactions secure. The sender commits to an amount of coins and this should remain private for the rest of the network over time.

\item \textit{Binding}: senders cannot change their commitments to a different transaction amount. If that were possible, it would mean that an adversary could spend coins which have already been committed to an UTXO, what would amount to create coins out of thin air.
\end{itemize}

There are two possible specifications for these properties. \textit{Computational hiding or binding} is when for all polynomial time adversaries, they can break the security property with negligible probability. This asymptotic security is parameterized by a security parameter $\lambda$ and adversaries run in polynomial time in $\lambda$ and their other inputs. On the other hand, we talk about \textit{perfectly hiding or binding}, when even with infinite computing power it would be not possible to break the security property.

Notice that a commitment scheme cannot be perfectly hiding and binding at the same time:

\begin{enumerate}[label=\roman*.]
\item If the scheme is perfectly hiding, there must exist several inputs committing to the same value. Otherwise, an adversary with infinite computing power attempting to find out which input committed to a certain output, could try all possible inputs finding out the corresponding output. This shows that this scheme cannot be perfectly binding.
\item If the scheme is perfectly binding, it means that there is at most one input that committed to an output. Imagine an adversary with infinite computing power attempting to find out which input committed to a target output. It would be possible to try all inputs and find which one verifies the commitment. Thus, this scheme cannot be perfectly hiding.
\end{enumerate}

So, for cryptocurrencies systems is better to provide stronger security in order to guarantee the hiding property. In other words, we prefer a commitment scheme with \textit{computational binding and perfectly hiding}. We can understand this by first assuming adversaries break the binding property. It means that they could create money from thin air from a certain point in time but this would not affect the blockchain history. On the other hand, if the adversary breaks the hiding property, history could be inspected and all the transactions revealed. That breaks one of the main principles of a privacy-oriented cryptocurrency.

\subsubsection{Pedersen commitments are computational binding} 
This property relies on the discrete logarithm assumption. In provable security, security is proved to hold against any probabilistic time adversary by showing an efficient way to break the cryptography protocol implies a way to break the underlying mathematical problem which is supposed to be hard (security reduction). The adversary is modeled as a procedure.
%\hlc{aca hay dos cosas. primero no esta claro que tiene que ver lo del pdersen commitment con lo que sigue. segundo, lo que sigue no esta bien redactado o no se entiende. no seria que se demuestra que el adversario puede romper el problema matematico en tiempo polinomial y esto implica que se puede romper el protocolo que esta construido sobre esa matematica? O sea no es al reves?}

%% se podria ver la prueba de reduccion

\begin{definition} [Computational Binding Commitment] 
\label{def:binding_commitment}

Let $\zeta (Setup, Com)$ be a Pedersen commitment scheme as in Definition \ref{def:pedersen_commitment}. Let $\adv_{Binding}$ be a polynomial probabilistic time adversary against the binding property running in the context of the game $G_{Binding}$ as in Figure \ref{game:game_bc}.
We say that the Pedersen commitment scheme $\zeta$ is computational binding if the success probability of $\adv_{Binding}$ winning game $G_{Binding}$ is negligible. 
\end{definition}

In game $G_{Binding}$, firstly the scheme is set up by choosing two generator points, $G$ and $H$, over the elliptic curve $\mathcal{C}$ of prime order $n$. All these parameters are public. Secondly, the adversary $\adv_{Binding}$ performs the attack attempting to find out two different transactional values $v_1$ and $v_2$ that commit to the same commitment. Once the adversary finishes the attack, two pair of different opening values are returned. The adversary succeeds if both pairs commit to the same value and $v_1 \neq v_2$.

\begin{figure}
\caption{Game Binding Commitment}
\label{game:game_bc}
\begin{center}
\framebox{
%\begin{minipage}{340pt}
\begin{tabular}{c}
$
\begin{array}{l}
\game{G_{Binding}} \eqdef\\
\quad \Call{(G,H,n)}{SetUp}{1^{\lambda}}\\
\quad \Call{(v_1,r_1), (v_2, r_2)}{\adv_{Binding}}{G,H,n};\\ 
\quad \Return Com(v_1,r_1) = Com(v_2,r_2)  \land $ $ v_1 \neq v_2 \\
\end{array}

$
\end{tabular}
%\end{minipage}
}
\end{center}
\end{figure}

As we mentioned before, the computational binding property is based on a security reduction. In terms of Pedersen commitment, it means that if the adversary $\adv_{Binding}$ could perform the attack in the context of game $G_{Binding}$ and could win with non-negligible probability, an adversary $\mathcal{I}_{Dlog}$ attacking a game against the discrete logarithm problem on the group $\mathcal{C}$ could use $\adv_{Binding}$ to win the game with non-negligible probability. 

Recall that MW uses Pedersen commitment with elliptic curves  (Definition \ref{def:pedersen_commitment}). The discrete logarithm problem on this context means: given a point $y$ over the elliptic curve $\mathcal{C}$ with generator $G$, it is hard to find $x$ such that $y=x.G$.

The following lemma captures the semantics of that security reduction. 

\begin{lemma} [Computational Binding] 
\label{lemma_binding_reduction} 
Let $\zeta (Setup, Com)$ be a Pedersen commitment scheme as in Definition \ref{def:pedersen_commitment}. 
Let $\adv_{Binding}$ be an adversary against the computational binding commitment (Definition \ref{def:binding_commitment}) in the commitment scheme $\zeta$.

Let us assume that $\adv_{Binding}$ succeeds in finding two distinct pair of opening values that commit to the same commitment with $\epsilon$ probability. 
Therefore, there exists an inversor $\mathcal{I}_{Dlog}$ which can find out the discrete logarithm to the base $G$ of a randomly chosen element $y$ on the elliptic curve $\mathcal{C}$ with $\epsilon'$ probability using the adversary $\adv_{Binding}$.

Hence, if $\epsilon'$ is negligible then $\epsilon$ is negligible too.

%$$
%\epsilon' \geq Simulation(\epsilon)
%$$
%$$
%t' \leq t + SimulationTime
%$$

\end{lemma}

In these cases, the contraposition is proved: if $\epsilon$ is non-negligible, then $\epsilon'$ is non-neglible too. 
The goal of the proof is to show how to transform the efficient adversary $\adv_{Binding}$ that is able to break the computational binding commitment into an algorithm that efficiently solves the discrete logarithm assumption. The inversor $\mathcal{I}_{Dlog}$ will provide a simulation context in which the adversary $\adv_{Binding}$ will perform its attack. The attack of the inversor $\mathcal{I}_{Dlog}$ will be successful if $\adv_{Binding}$ is successful and the simulation does not fail.

%The challenge of $\mathcal{I}$ is to find the discrete logarithm of a randomly choosen element $y$, so $\mathcal{I}_{Dlog}$ will provide $y$ to $\adv_{Binding}$ in an \textit{intelligent} way. The attack of the inversor $\mathcal{I}_{Dlog}$ will be successful if $\adv_{Binding}$ is successful and the simulation does not fail. 

According to game $G_{Binding}$, when the adversay $\adv$ succeeds we have two identical commitments $Com(v_1,r_1) = Com(v_2,r_2)$ and $v_1 \neq v_2$ such that (Definition \ref{def:pedersen_commitment}):

$$
r_1 . G + v_1 . H = r_2 . G + v_2 . H
$$ 
So we can compute:
$$
H = \dfrac{r_1 - r_2}{v_2 - v_1}.G
$$
Which means that we have computed the discrete logarithm of $H$ with respect to $G$. 
%It breaks our security assumption of the scheme about that the discrete logarithm relative to each other were unknown.

Figure \ref{game:game_inversor} shows the game $G_{Dlog}$ which captures the semantic of the reduction. The failure event captures when the adversary $\adv_{Binding}$ fails and therefore, the adversary $\mathcal{I}_{Dlog}$ fails too.

The probability of success of $\mathcal{I}_{Dlog}$ is equal to the probability of success of $\adv_{Binding}$.

\begin{figure}
\caption{Game Inversor DLog}
\label{game:game_inversor}
\begin{center}
\framebox{
%\begin{minipage}{340pt}
\begin{tabular}{cc}
$
\begin{array}{l}
\game{G_{Dlog}} \eqdef\\
\quad \Call{(G,H,n)}{SetUp}{1^{\lambda}}\\
\quad \Call{y}{\mathcal{I}_{Dlog}}{G,H,n}\\ 
\quad \mathsf{if}\ {y = failure}\ \mathsf{then}\\ 
\qquad \Return failure \\
\quad \mathsf{else} \\
\qquad \Return H = y.G 
\end{array}
\quad
$
&
$
\begin{array}{l}
\mathcal{I}_{Dlog}(G,H,n) \eqdef\\
\quad \Call{(v_1,r_1), (v_2, r_2)}{\adv_{Binding}}{G,H,n}\\ 
\quad \mathsf{if}\ {Com(v_1,r_1) = Com(v_2,r_2)  \land v_1 \neq v_2} $ $ \mathsf{then}\\
\qquad \Return \dfrac{r_2 - r_1}{v_2 - v_1} \\
\quad \mathsf{else} \\
\qquad \Return failure
\end{array}
$
\end{tabular}
%\end{minipage}
}
\end{center}
\end{figure}

\subsubsection{Pedersen commitments are perfectly hiding} 
Basically, it is because, given a commitment $Com(v,r)=r.G + v.H$, there are many combinations of $(v',r')$ that satisfies $Com(v,r)=r'.G + v'.H$. 
Despite the adversary have infinite computing power and could attempt all possible values, there would be no way to know which opening values $(v',r')$ were the original ones. Furthermore, $r$ is a random value of the finite field $\mathbb{F}_n$ so $r.G + v.H$ is a random element of $\mathcal{C}$.  
%independent of the choice of v

\begin{definition} [Perfectly Hiding Commitment] 
\label{def:hiding_commitment}
Let $\zeta (Setup, Com)$ be a Pedersen commitment schema as in Definition \ref{def:pedersen_commitment}. Let $\adv$ be a computationally
unbounded adversary against the hiding property running in the context of the game $G_{Binding}$ as in Figure \ref{game:game_hc}.
We say that the Pedersen commitment scheme $\zeta$ is perfectly binding if the success probability of $\adv$ winning game $G_{Hiding}$ holds:
$$
Pr ( b = b' ) = \dfrac{1}{2}
$$
\end{definition}

\begin{figure}
\caption{Game Hiding Commitment}
\label{game:game_hc}
\begin{center}
\framebox{
%\begin{minipage}{340pt}
\begin{tabular}{c}
$
\begin{array}{l}
\game{G_{Hiding}} \eqdef\\
\quad \Call{(G,H,n)}{SetUp}{1^{\lambda}}\\
\quad \Call{(v_0,v_1)}{\adv_{Hiding}}{G,H,n}\\
\quad \Rand{b}{\{0,1\}}\\
\quad \Rand{r}{\mathbb{F}_n}\\
\quad com = Com(v_b,r) \\
\quad \Call{b'}{\adv}{com,G,H,n}\\
\quad \Return b = b' \\
\end{array}

$
\end{tabular}
%\end{minipage}
}
\end{center}
\end{figure}

In the game described in Figure \ref{game:game_hc}, first the game is set up  and then the adversary chooses two distinct transactional values $v_0$ and $v_1$.
Then, one of these values is randomly choosen as $v_b$, as well as with the blinding factor $r$. The commitment of $(v_b,r)$ is computed and the adversary $\adv_{Hiding}$ performs the attack attempting to find out which one of the values was committed.

\subsubsection{Switch commitments}
As already mentioned, if an attacker succeeds in breaking the computational binding property of a commitment then money can be created from thin air.
Switch commitments \cite{cryptoeprint:2017:237} were introduced to enable the transition from \textit{computational bindingness} to \textit{statistical bindingness}, specially to the commitments stored in the blockchain. The notion of statistical security implies that a computationally unbounded adversary cannot violate the property except with negligible probability.

If in a certain moment we believe that the bindingness of the commitment scheme gets broken, we could make a soft fork on the chain and switch existing commitments to this new validation scheme which is backwards compatible.

Below, we show the changes that are needed for our model to also encompass Switch commitments. In Pedersen commitment definition (Definition \ref{def:pedersen_commitment}) we add a third point generator $J$ of the elliptic curve $\mathcal{C}$ whose discrete logarithm relative to $G$ and $H$ is unknown.
We define the new commitment algorithm as follows:

$$
Com(v,r) = r'.G + v.H, \textit{with } v~ \textit{the transactional value and} 
$$
$$
r' = r + hash( v.H + r.G  ,  r.J ), 
$$
% no sé, me parece que queda más claro que parte se sustituye por otra, pero se puede cambiar
%\hlc{por que no escribir $Com(v,r) = (r + hash( vH + rG  ,  rJ ))G + vH$?}
where $r$ is the blinding factor randomly chosen in the finite field $\mathbb{F}_n$ and
$J$ is the third point generator.

Note that $r$ is still randomly distributed and the hash value of \textit{ElGamal} commitment is computed which is the combination of ElGamal encryption \cite{crypto-1984-1321} and a commitment scheme. 

%A commitment from collision-resistant hash functions was proposed by Shai Halevi and Silvio Micali [HM96]

%\paragraph{Zero-knowledge Proof.}
%\noindent
%\paragraph{{\bf Zero-knowledge Proof}}
\subsection{Zero-knowledge Proof}
%\textbf{Zero-knowledge Proof.}
The goal is to prove that a statement is true without revealing any information beyond the verification of the statement. 
In MW  we need to ensure that in every transaction the amount is positive so that users cannot create coins. Here, the hard part is to prove that without revealing the amount. In our model, the output amounts are hidden in the form of a Pedersen commitment, and the transaction contains a list of range proofs of the outputs to prove that the amount is positive. MW uses \textit{Bulletproofs} %\cite{BunzBullet} 
to achieve this goal. In our model, this verification is performed as the first step of the validation of the transaction. %If we refer to our model, we pointed out this check as first step of the validity of a transaction.
%For that reason, we need to specify a verifier algorithm that takes a list of range proofs and performs the verification. 
%\gusnote{Este parrafo está muy confuso!} \hlc{maxi: yo mas bien no logro entender la relacion entre que las range proof sean validas y la necesidad de definir un algoritmo que haga esa verificacion}

%The proof essentially opens the commitment to the homomorphic sum of the inputs minus the outputs to zero, which does not reveal the individual monetary amounts of the inputs and out- puts in the transaction. To be sound, a non-interactive zero-knowledge proof is added to each commitment to show that the committed value is in a certain range. These so-called range proofs ensure that the computation of the sum does not overflow.

%\paragraph{{\bf Unlinkability}}
\subsection{Unlinkability and Untraceability}
\label{secc:unlinkability} 
MW does not use addresses; the protocol relies on confidential transactions to hide the identity of the sender and the recipient. It means that users have to communicate off-chain to create the transactions. 

As we specified in our model, each node has a pool of unconfirmed transactions in the \textit{mempool}. This transactions are waiting for the miners in order to be included in a block. We can distinguish two security properties of the transactions. Untraceability refers to the transactions in the mempool and unlinkability to the transactions in the block. In our model, this two notions are formalized as follows.

\begin{property}[Transaction Unlinkability]
\label{prop_unlinkability} 
Given a valid block $b$, it is computationally infeasible to know which input cancels which output.
\end{property}

The following lemma captures the semantics of this property in MW. Moreover, the operations cut-through and CoinJoin, which were described above, also contribute to this property. 

%\begin{Property}[Transaction Unlinkability]
%\label{prop_unlinkability} 
%Given a valid block $b$, it is computationally infeasible to know which input cancels which output.
%\end{Property}

\begin{lemma} [Transaction Unlinkability] 
\label{lemma_unlinkability} 
%It is said a valid block $b$ is transaction-unlinkable if for any polynomial probabilistic time adversary $\adv$, the probability of finding a balanced transaction within the block is negligible.
For any valid block $b$ and for any polynomial probabilistic time adversary $\adv$, the probability of $\adv$ in finding a balanced transaction within $b$ is negligible.
%\hlc{diculpen pero para mi esto sigue estando confuso porque mezcla una definicion con una tesis. Esto `we say block $b$ is transaction-unlinkable' es defincion. Para mi el lema es: For any valid block $b$ and for any polynomial probabilistic time adversary $\adv$, the probability of $\adv$ in finding a balanced transaction within $b$ is negligible. Luego o antes podemos definir: Definition [Transaction Unlinkability] We say block $b$ is transaction-unlinkable, if the probability of any $\adv$ in finding a balanced transaction within $b$ is negligible. Y luego podemos decir: Due to Lemma 7, in MW all blocks are transaction-unlinkeable.}
\end{lemma}

%\begin{proof}
\subsubsection*{{\bf Proof.}} 
Let $b= \{i, o, tks, ko\}$ be a valid block with $tks = (tk_1, ..., tk_t)$ the list of transaction kernels. The $j$-th item in $tks$ is of the form $tk_j = \{rp_j, ke_j, \sigma_j\}$. 

The goal of the adversary $\adv$ is to find a tuple of the form $\{i',o',ke'\}$ where the list of inputs $i'$ is a subset of $i$ and the list of outputs $o'$ is a subset of $o$, satisfying Definition \ref{def:balanced_transac} of a balanced transaction.
It means that, the following equality must be true for the tuple:

$$
\sum_{o_j \in o'}{o_j} - \sum_{c_j \in i'}{c_j} = ke' + tko'.G
$$
where $ke'$ is the transaction excess and $tko'$ the transaction kernel offset.

If we refer to the construction process in Definition \ref{def:aggregate_block}, the transaction kernel offsets were added to generate a single aggregate offset $ko$ to cover all transactions in the block. It means that we do not store the individual kernel offset $tko'$ of the transaction in $b$ once the transaction is aggregated to the block.

The challenge is trying to solve the adversary $\adv$ could be seen as the subset sum problem (\textit{NP-complete}) but, in this case, $tko'$ is unrecoverable. So, although many transactions have few inputs and outputs, it is computationally infeasible, without knowing that value, to find the tuple.
\qed
%\end{proof}

We can define:

\begin{definition}[Transaction Unlinkable]
\label{def:unlinkability} 
We say block $b$ is transaction-unlinkable, if the probability of any polynomial probabilistic time adversary $\adv$ in finding a balanced transaction within $b$ is negligible. 
\end{definition}

Then, due to Lemma \ref{lemma_unlinkability} we conclude that in MW all blocks are transaction-unlinkeable.

\begin{property}[Transaction Untraceability]
\label{prop_untraceability} 
For every transaction in the mempool, it is not possible to relate the transaction to the IP address of the node which originated it.
\end{property}

Regarding this property, we should refer mainly, to the broadcast of the transactions. Once the transactions are created, they are broadcasted to the network and they are included in the mempool. Each node could track the IP address from the node which received the transaction. At that point nodes could record the transactions, allowing them to build a transaction graph. 

We define that the broadcast of a transaction can be performed with or without confusion. 
Without confusion means that, once the transactions are created, they are broadcasted immediately to all the network. However, if someone controls enough nodes on the network and discovers how the transaction moves, he could find out the IP address node from which the transaction comes from. 
%The success of this attack is between 11% to 60% -> cite: Deanonymisation of clients in Bitcoin P2P network (no la agregue)

On the other hand, we define the broadcast with confusion as a way to obscure the IP address node. 

\begin{definition}[Broadcast with confusion]
\label{prop:broadcast_conf}
Let's say node $A$ sends a transaction to node $B$. We say $B$ receives the transaction with confusion if given the IP address of node $A$, the node $B$ does not know if the transaction was originated by the node $A$ or not.
\end{definition}
% ver de poner lo de probabilidad o no: the node does not know, with enough probability, if the transaction

In other words, it can be said that if some malicious nodes, working together, construct a graph of the pairs $(\mathit{transaction}, \mathit{IP\ address\ node})$, the IP address node will not convey information about what node originated the transaction. Therefore, in our model, we require Property \ref{prop:broadcast_conf} to hold before the broadcast takes place. In order to achieve this, we can establish that the node broadcasting the transaction should be far enough from the one which originated it. Moreover, CoinJoin could be performed before the broadcast. 

Dandelion, proposed by Fanti et al \cite{DBLP:journals/corr/Venkatakrishnan17}, is a protocol for transaction broadcasting intended to resist that deanonymization attack. 
Dandelion is not part of the MW protocol, however this kind of protocols should be implemented by each node to lower the risk of creating the transaction graph. In Dandelion, broadcasting is performed in two phases: the ``steam'' phase and the ``fluff'' phase. In the ``steam'' phase the transaction is broadcasted randomly to one node, which then randomly sends it to another, and so on. This process finishes when the ``fluff'' phase is reached, and the transaction is broadcasted to the network. 
The following routines capture the semantic of the phases: 
\vspace{2ex}

$\begin{array}{l}
subroutine\ steam(tx:Transaction) \{\\
c \leftarrow \{0,1\}~\text{(* random decision *)} \\
if\ c == 0 \ then \\
~~~~~node \leftarrow select\_random\_node() \\
~~~~~node.steam(tx) \\
else \\
~~~~~this.fluff(tx) \\	
\}\\ \\
\end{array}$

$\begin{array}{l}
subroutine\ fluff(tx:Transaction) \{\\
     broadcast(tx) \\
\}\\ \\
\end{array}$

Each node, besides having the local state, should implement these two routines. Once the transaction is created and is ready to be included in the mempool, its broadcasting start in the ``steam'' phase. When it reaches the ``fluff'' phase, it is broadcasted to the network and added in the mempool.

Dandelion relies on the following three rules: all nodes obey the protocol, each node generates exactly one transaction, and all nodes on the network run Dandelion. 
%\hl{que diferencia hay entre que cada nodo obedece el protocolo y que cada nodo ejecuta Dandelion? No es Dandelion el protocolo?}
% en varios lados habla como si fuera diferente, habría que leer bien el paper
The problem is that an adversary can violate them. For that reason, \texttt{Grin} implements a more advanced protocol called Dandelion++ \cite{DBLP:journals/corr/abs-1805-11060} which intends to prevent that \cite{dandelion_plus_plus_grin}. However, it is believed that Dandelion++ is not good enough to guarantee the privacy of a virtual coin \cite{grin_privacy_primer}. For instance, the flashlight attack \cite{blockchain_privacy} is an open problem still under investigation \cite{open_problems_grin}. The scenario here is when an `activist' want to accept donations but he cannot reveal his identity. At some point, he will deposit those payments to an exchange and his identity would be compromised. The adversary injects `tainted coins' and could build a `taint tree' looking through all deposits to the exchange. This way, he could link those deposits to the `activist'. 

%Some approaches were presented as a countermeasure of such attack. For instance, the combination of the MW protocol with a Zerocash-style commitment-nullifier scheme \cite{ethereum_zk_snark}. On the other hand, it could be useful to analyse how other virtual coins are resistant to it. In the case of \textit{Zcash}, every shielded transaction has a large anonymity set (the set of transactions that the transaction is indistinguishable from). In the case of Spectrecoin \cite{spectre_coin_white_paper} the main idea is the use, only once, of public addresses (XSPEC) to receive the payments combined with an anonymous staking protocol.

The combined use of the MW protocol with a Zerocash-style commitment-nullifier schema has been put forward in \cite{ethereum_zk_snark} as a countermeasure to the above attack. In the case of \textit{Zcash}, every shielded transaction has a large anonymity set, namely, a set of transactions form which it is indistinguishable from. In the case of Spectrecoin \cite{spectre_coin_white_paper} the main idea is the use, only once, of public addresses (XSPEC) to receive the payments combined with an anonymous staking protocol.

%\paragraph{\bf Model-driven verification}
%\hlc{cambie a subsubsection}
%\noindent
%\paragraph{{\bf Model-driven verification}}
\section{Model-driven verification}
\label{sec:mdd}
%\textbf{Model-driven verification.}
MW is built on top of a consensus protocol. 
%Hence, in part, analyzing MimbleWimble implies the analysis of a consensus protocol.
In that direction, we have developed a Z specification of one such protocol, part of which is included in what follows. Z specifications in turn can be easily translated into the \setlog language \cite{DBLP:journals/jar/CristiaR20}, which can be used both as a (prototyping) programming language and a satisfiability solver for an expressive fragment of set theory and set relation algebra. We present an excerpt of the \setlog prototype after its Z specification.  This \setlog prototype can be used as an executable model where simulations can be run. This allows us to analyze the behavior of the protocol without having to implement it in a low level programming language.
%
%\hlc{The consensus protocol isn't MimbleWimble. Is this a problem? Should we say that we can easily develop a Z specification and a setlog prototype of it?}

%\subsection{Formal Proof}
We also plan to use \setlog to prove some of the basic properties mentioned above, such as the invariance of $valid\_state$. However, for complex properties or for properties not expressible in the set theories supported by \setlog we plan to develop a complete and uniform formulation of several security properties of the protocol using the \texttt{Coq} proof assistant \cite{coq-manual}. 
The \texttt{Coq} environment supports advanced logical and
computational notations, proof search and automation, and modular development of
theories and code. It also provides program extraction towards languages like
\texttt{Ocaml} and \texttt{Haskell} for execution of (certified) algorithms \cite{conf/types/Letouzey02}. Additionally,
\texttt{Coq} has an important set of libraries; for example \cite{DBLP:conf/itp/BartziaS14} contains a formalization of elliptic curves theory, which allows the verification of elliptic curve cryptographic algorithms. 
%\hlc{new} 

The fact of first having a \setlog prototype over which some verification activities can be carried out without much effort helps in simplifying the process of writing a detailed \texttt{Coq} specification. This is in accordance with proposals such as \texttt{QuickChick} whose goal is to decrease the number of failed proof attempts in \texttt{Coq} by generating counterexamples before a proof is attempted \cite{denes2014quickchick}.

%\subsection{Model-Based Testing}
%\hlc{moved from below}
By applying the program extraction mechanism provided by \texttt{Coq} we would be able  to derive a certified \texttt{Haskell} prototype of the protocol. 
This prototype can be used as a testing oracle and also to conduct further verification activities on correct-by-construction implementations of the protocol. In particular, both the \setlog and \texttt{Coq} approaches can be used as forms of model-based testing. That is, we can use either specification to automatically generate test cases with which protocol implementations can be tested \cite{DBLP:conf/sefm/CristiaRF13,denes2014quickchick}.

\subsection{\label{zModel}Excerpt of a Z model of a consensus protocol}
The following is part of a Z model of a consensus protocol based on the model developed by P\^irlea and Sergey \cite{Pirlea:2018:MBC:3176245.3167086}. For readers unfamiliar with the Z notation we have included some background in Appendix \ref{backz}.

The time stamps used in the protocol are modeled as natural numbers. Then we have the type of addresses ($Addr$), the type of hashes ($Hash$), the type of proofs objects ($Proof$) and the type of transactions ($Tx$). Differently from P\^irlea and Sergey's model\footnote{From now on we will refer to P\^irlea and Sergey model simply as PS.} we model addresses as a given type instead as natural numbers. In PS the only condition required for these types is that they come equipped with equality, which is the case in Z.
\begin{zed}
Time == \nat \also
[Addr,Hash,Proof,Tx]
\end{zed}
The block data structure is a record with three fields: $prev$, (usually) points to the parent block; $txs$, stores the sequence of transactions stored in the block; and $pf$ is a proof object required to validate the block.
\begin{schema}{Block}
prev:Hash \\
txs: \seq Tx \\
pf:Proof
\end{schema}

Next we define the following parameters of the model: $hashb$, a function computing the hash of a block; $hasht$, a function computing the hash of a transaction; $mkProof$, a function computing a proof object of a node; $VAF$, a relation used to validate proof objects; $txValid$, a relation used to validate transactions; and $txExtend$, a relation used to modify a set
of pending transactions stored in a node. $VAF$ is constrained to triplets where the block whose proof is being considered is not one of the blocks of the chain being considered.

\begin{axdef}
hashb:Block \inj Hash \\
hasht:Tx \inj Hash \\
mkProof:Addr \cross \seq Block \pfun Proof \\
VAF: \power(Proof \cross Time \cross \seq Block) \\
txValid:Tx \rel \seq Block \\
txExtend: \power Tx \cross Tx \fun \power Tx
\where
\forall b:Block; t:Time; c:\seq Block @ (b.pf,t,c) \in VAF \implies b \notin \ran c
\end{axdef}

Another parameter of our model is the \emph{genesis block}, called $GB$, which should be provided by the client of the model. Clearly, $GB$ is a block enjoining two particular properties: it has no parent and it contains no transactions. 

\begin{axdef}
GB:Block
\where
GB.prev = hashb~GB \\
GB.txs = \langle \rangle
\end{axdef}

The local state space of a participating network node is given by four state variables: $this$, representing the address of the node; $as$, are the addresses of the peers this node is aware of; $bf$, is a block forest which records the minted and received blocks; and $tp$, is a set of received transactions which eventually will be included in minted blocks.
\begin{schema}{LocState}
this:Addr \\
as:\power Addr \\
bf: Hash \pfun Block \\
tp: \power Tx
\end{schema}

As nodes can send messages we define their type. $NullMsg$ is used when, actually, the node does not send any message (think of it as the null statement in programming languages, i.e. $skip$). Also note that some messages contain some data, e.g. $AddrMsg$ which communicates a set of peers by transmitting their addresses.

\begin{zed}
Msg ::= \\
  \t1 NullMsg \\
  \t1 | ConnectMsg \\
  \t1 | AddrMsg \ldata \power Addr \rdata \\
  \t1 | TxMsg \ldata Tx \rdata \\
  \t1 | BlockMsg \ldata Block \rdata \\
  \t1 | InvMsg \ldata \power Hash \rdata \\
  \t1 | GetDataMsg \ldata Hash \rdata
\end{zed}

Packets are used to build so-called \emph{packet soups} which are used later to define the system configuration (see schema $Conf$). A packet is a triple where the first
component is the message sender, the second is the
destination's address and the third is the
message content.

\begin{zed}
Packet == Addr \cross Addr \cross Msg
\end{zed}

The model has twelve state transitions divided into two groups: \emph{local} and \emph{global}. Local transitions are those executed by network nodes, while global transitions promote local transitions to the network level. In turn, the local transitions are grouped into \emph{receiving} and \emph{internal} transitions. Receiving transitions model the nodes receiving messages from other nodes and, possibly, sending out new messages; internal transitions model the execution of instructions run by each node when some local condition is met. 

With the model elements defined so far we can give the specification of the local transitions. We start with $RcvNull$ which models a rather trivial operation of the protocol when actually the state of the node does not change because the $NullMsg$ has been received.

\begin{schema}{RcvNull}
\Xi LocState \\
p?:Packet \\
ps!:\power Packet
\where
p?.2 = this \\
p?.3 = NullMsg \\
ps! = \emptyset
\end{schema}

$RcvConnect$ specifies the transition where the node receives a $ConnectMsg$ message making it to add the sender's address as a new peer. In this transition we see that a node can output a set of packets (which in this particular case is a singleton set) as a side effect of receiving a message. In this case the packet says that $this$ node is sending the packet addressed to the node that just sent a packet to $this$. In turn the payload of the packet is an $InvMsg$ message which informs the destination the transactions and blocks stored by $this$.

\begin{schema}{RcvConnect}
\Delta LocState \\
p?:Packet \\
ps!:\power Packet
\where
p?.2 = this \\
p?.3 = ConnectMsg \\
as' = as \cup \{p?.1\} \\
this' = this \\
bf' = bf \\
tp' = tp \\
ps! = \{(this,p?.1,InvMsg (\dom bf \cup hasht\limg tp \rimg))\}
\end{schema}

The next transition is $RcvAddr$. As can be seen, it sends out a set of packets which can potentially have many elements. The node checks whether or not the packet's destination address coincides with its own address. In that case, the node adds the received addresses to its local state and sends out a set of packets that are either of the form $(this,a,ConnectMsg)$ or $(this,a,AddrMsg~as')$. The former are packets generated from the received addresses and sent to the new peers the node now knows, while the latter are messages telling its already known peers that it has learned of new peers.
\begin{schema}{RcvAddr}
\Delta LocState \\
p?:Packet \\
ps!:\power Packet
\where
p?.2 = this \\
\exists asm:\power Addr @ \\
   \t1 p?.3 = AddrMsg~asm \\
   \t1 \land as' = as \cup asm \\
   \t1 \land this' = this \\
   \t1 \land bf' = bf \\
   \t1 \land tp' = tp \\
   \t1 \land ps! = \\
   \t2 \{a:asm \setminus as @ (this,a,ConnectMsg)\} \\
   \t2{} \cup \{a:as @ (this,a,AddrMsg~as')\}
\end{schema}

$RcvTx$ specifies the reception of a new transaction by a node.
In this case the node adds the transaction to
its local state and sends out an $InvMsg$ to its peers telling them that $this$ is now possessing the received transaction.

\begin{schema}{RcvTx}
\Delta LocState \\
p?:Packet \\
ps!:\power Packet
\where
p?.2 = this \\
\exists tx:Tx @ \\
   \t1 p?.3 = TxMsg~tx \\
   \t1 \land this' = this \\
   \t1 \land as' = as \\
   \t1 \land bf' = bf \\
   \t1 \land tp' = txExtend(tp,tx) \\
   \t1 \land ps! = \{a:as @ (this,a,InvMsg(\dom bf \cup hasht \limg tp' \rimg))\}
\end{schema}

The next transition is $RcvBlock$ which specifies a node receiving a block instead of a transaction. However, in order to specify $RcvBlock$ we first need to introduce several elements in the form of parameters to the model. We start by introducing $FCR$ as an order relation on the set of block chains. Note that the axioms imply that $FCR$ is total, transitive and irreflexive. The fourth axiom states that extensions of a chain are ``heavier'' than the chain itself.

\begin{axdef}
FCR: \seq Block \rel \seq Block
\where
\forall x,y:\seq Block @ (x,y) \in FCR \lor (y,x) \in FCR \lor x = y \\
\forall x,y,z:\seq Block @ (x,y) \in FCR \land (y,z) \in FCR \implies (x,z) \in FCR \\
\forall x:\seq Block @ \lnot (x,x) \in FCR \\
\forall x,y:\seq Block; b:Block @ (x,x \cat y \cat \langle b \rangle) \in FCR \\
\forall x,y:\seq Block @ x \prefix y \implies (x,y) \in FCR \lor x = y
\end{axdef}

The function $maxFCR$ returns the maximum chain of a set of chains according to the $FCR$ order.

\begin{axdef}
maxFCR:\power(\seq Block) \fun \seq Block
\where
\forall C:\power(\seq Block); m:\seq Block @ \\
  \t1 maxFCR~C = m \iff m \in FCR \limg C \rimg \land m \notin \dom(C \dres FCR)
\end{axdef}

Function $chain$ is one of the building blocks necessary to compute the ledger of a block forest. Block forests are represented as partial functions from hashes to blocks (formally $Hash \pfun Block$). Then, $chain$ takes as inputs a block forest and a block an returns a block chain. We will use $chain$ to ``iterate'' over all the blocks of a given block forest.

\begin{axdef}
chain:(Hash \pfun Block) \fun Block \fun \seq Block
\where
\forall bf:Hash \pfun Block; b:Block; c:\seq Block @ \\
  \t1 chain~bf~b = c \iff \\
%    \t2 \ran c \subseteq \ran bf \\
    \t2  c \in \nat_1 \inj \ran bf \\
    \t2 \land head~c = GB \\
    \t2 \land (b \notin \ran bf \implies c = \langle GB \rangle) \\
    \t2 \land (b \in \ran bf \implies last~c = b) \\
    \t2 \land (\forall i:\dom c @ 
                  i > 1 \implies (c~i).prev = hashb(c(i-1))) \\
    \t2 \land (\forall x:\ran c @ \\
      \t3       (\ran x.txs) \cross 
                \{y:\seq Block | y \prefix c \land x \notin \ran y\}
                  \subseteq txValid)
\end{axdef}

Now we define the ledger of a block forest as the longest block chain returned by $chain$ for each block in the forest.

\begin{axdef}
ledger:(Hash \pfun Block) \fun \seq Block
\where
\forall bf:Hash \pfun Block @ ledger~bf = maxFCR \{b:\ran bf @ chain~bf~b\} 
\end{axdef}

Finally, we can give the specification of $RcvBlock$.

\begin{schema}{RcvBlock}
\Delta LocState \\
p?:Packet \\
ps!:\power Packet
\where
p?.2 = this \\
\exists b:Block @ \\
   \t1 p?.3 = BlockMsg~b \\
   \t1 \land this' = this \\
   \t1 \land as' = as \\
   \t1 \land bf' = bf \oplus \{hashb~b \mapsto b\} \\
   \t1 \land tp' = \dom(tp \dres txValid \rres \{ledger(bf \oplus \{hashb~b \mapsto b\})\})  \\
   \t1 \land ps! = \{a:as @ (this,a,InvMsg(\dom bf' \cup hasht \limg tp' \rimg))\}
\end{schema}

As can be seen, $RcvBlock$ adds the received block to the block forest without checking its validity. This is so because the node might not have received the preceding blocks which determine the validity of the received block.

$RcvInv$ specifies the behavior of the node when it receives an $InvMsg$ message. Such a message is used to inform nodes of the transactions and blocks stored by a given node. Then, when $this$ receives an $InvMsg$ message it asks the system the transactions and block it does not know yet by means of a $GetDataMsg$ message.

\begin{schema}{RcvInv}
\Xi LocState \\
p?:Packet \\
ps!:\power Packet
\where
p?.2 = this \\
\exists hs:\power Hash @ \\
   \t1 p?.3 = InvMsg~hs \\
   \t1 \land ps! = \{h:hs \setminus (\dom bf \cup hasht \limg tp \rimg) @ (this,p?.1,GetDataMsg~h)\}
\end{schema}

The last receiving local transition is $RcvGetData$. This operation is divided into three cases: the node receiving a block ($RcvGetDataBlock$); the node receiving a transaction ($RcvGetDataTx$); and the node receiving data it has not requested ($RcvGetDataNull$). 

\begin{schema}{RcvGetDataBlock}
\Xi LocState \\
p?:Packet \\
ps!:\power Packet
\where
p?.2 = this \\
\exists h: Hash @ \\
   \t1 p?.3 = GetDataMsg~h \\
   \t1 \land h \in \dom bf \\
   \t1 \land ps! = \{(this,p?.1,BlockMsg(bf~h))\}
\end{schema}

\begin{schema}{RcvGetDataTx}
\Xi LocState \\
p?:Packet \\
ps!:\power Packet
\where
p?.2 = this \\
\exists h: Hash @ \\
   \t1 p?.3 = GetDataMsg~h \\
   \t1 \land h \in \ran(tp \dres hasht) \\
   \t1 \land ps! = \{(this,p?.1,TxMsg((tp \dres hasht)\inv h))\}
\end{schema}

\begin{schema}{RcvGetDataNull}
\Xi LocState \\
p?:Packet \\
ps!:\power Packet
\where
p?.2 = this \\
\exists h: Hash @ \\
   \t1 p?.3 = GetDataMsg~h \\
   \t1 \land h \notin \ran(tp \dres hasht) \\
   \t1 \land h \notin \dom bf \\
   \t1 \land ps! = \{(this,p?.1,NullMsg)\}
\end{schema}

\begin{zed}
RcvGetData \defs RcvGetDataBlock \lor RcvGetDataTx \lor RcvGetDataNull
\end{zed}

\subsection{\label{setlogModel}Excerpt of the \setlog prototype generated from the Z specification}
In this section we show part of the \setlog code corresponding to the Z model presented above. \setlog code can be seen as both a formula and a program \cite{DBLP:journals/jar/CristiaR20}. Thus, in this case we use the code as a prototype or executable model of the Z model. The intention is twofold: to show that passing from a Z specification to a \setlog program is rather easy, and to show how a \setlog program can be used as a prototype. The first point is achieved mainly because \setlog provides the usual Boolean conectives and most of the set and relational operators available in Z. Hence, it is quite natural to encode a Z specification as a \setlog program.

Given that \setlog is based on Prolog its programs resemble Prolog programs. The \setlog encoding of $RcvAddr$ is the following:
\begin{verbatim}
rcvAddr(LocState,P,Ps,LocState_) :-
LocState = {[as,As], [this,This] / Rest} &
P = [_,This, addrMsg(Asm)] & un(As,Asm,As_) & 
diff(Asm,As,D) &
Ps1 = ris(A in D,[],true,[This,A,connectMsg]) &
Ps2 = ris(A in As,[],true,[This,A,addrMsg(As_)]) & 
un(Ps1,Ps2,Ps) &
LocState_ = {[as,As_], [this,This] / Rest}.
\end{verbatim}

As can be seen, \verb+rcvAddr+ is clause receiving the before state (\verb+LocState+), the input variable (\verb+P+), the output variable (\verb+Ps+) and the after state (\verb+LocState_+). As in Prolog, \setlog programs are based on unification with the addition of set unification. In this sense, a statement such as \verb+LocState = {[as,As], [this,This] / Rest}+ (set) unifies the parameter received with a set term singling out the state variables needed in this case (\verb+As+ and \verb+This+) and the rest of the variables (\verb+Rest+). The same is done with packet \verb+P+ where \verb+_+ means any value as first component and \verb+addrMsg(Asm)+ gets the set of addresses received in the packet without explicitly introducing an existential quantifier.

The set comprehensions used in the Z specification are implemented with \setlog's so-called Restricted Intentional Sets (RIS) \cite{DBLP:conf/cade/CristiaR17}. A RIS is interpreted as a set comprehension where the control variable ranges over a finite set (\verb+D+ and \verb+As+). 

Given \verb+rcvAddr+ we can perform simulations on \setlog such as:
\begin{verbatim}
S = {[as,{}], [this,This] / R} & 
rcvAddr(S,[_,This,addrMsg({a1,a2})],P1,S1) & 
rcvAddr(S1,[_,This,addrMsg({a1,a3})],P2,S2).
\end{verbatim}
in which case \setlog returns:
\begin{verbatim}
P1 = ris(A in {a1,a2/_N2},[],true,[This,A,connectMsg],true),  
S1 = {[as,{a1,a2}], [this,This] / R},  
P2 = {[This,a3,connectMsg],[This,a1,addrMsg({a2,a1,a3})],
      [This,a2,addrMsg({a2,a1,a3})] /
      ris(A in _N1,[],true,[This,A,connectMsg],true)},  
S2 = {[as,{a2,a1,a3}], [this,This] / R}
Constraint: subset(_N2,{a1,a2}), subset(_N1,{a1,a3}), 
            a1 nin _N1, a2 nin _N1
\end{verbatim}
That is, \setlog binds values for all the free variables in a way that the formula is satisfied (if it is satisfiable at all). In this way we can trace the execution of the protocol w.r.t. states and outputs by starting from a given state (e.g. \verb+S+) and input values (e.g. \verb+[_,This,addrMsg({a1,a2})]+), and chaining states throughout the execution of the state transitions included in the simulation (e.g. \verb+S1+ and \verb+S2+).

\section{Mimblewimble implementations}
\label{sec:implementations}

In August 2016, someone called "Tom Elvis Jedusor'' (french name for Voldemort in Harry Potter) posted a link to a text file on the IRC Channel describing a cryptocurrency protocol with a different approach from BitCoin. This article titled `Mimblewimble' \cite{mimbleimble-wp} addressed some privacy concerns and the ability of compressing the transaction history of the chain without loss of validity verification.
Since this document left some questions open, in October 2016 Andrew Poelstra published a paper \cite{poelstra16} where he describes, in more detail, the design of a blockchain based on Mimblewible.
In 2019, the first two practical implementations were launched: \texttt{Grin} and \texttt{Beam}.  In what follows, we shall first describe the main features of their design and will compare them with our model. Then, we shall discuss features that set apart \texttt{Grin} from \texttt{Beam}.

\subsection{Grin}
\texttt{Grin} \cite{mimbleimble-doc} is an open source software project with a simple approach to MW. As we will see below, its design is a straightforward interpretation of our model. 

\subsubsection{Blocks and Transactions}
\begin{figure}
	\centering
	\includegraphics[width=0.65\textwidth]{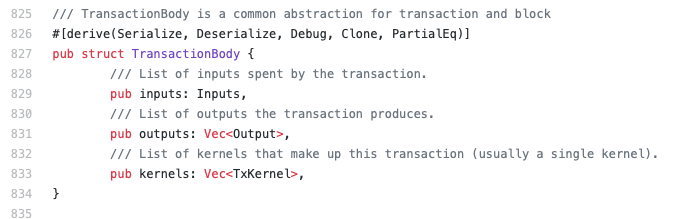}
	\caption{\texttt{Grin} transaction body source code \cite{grin_github}}
	\label{fig:grin_transac} 
\end{figure}

\begin{figure}
	\centering
	\includegraphics[width=0.6\textwidth]{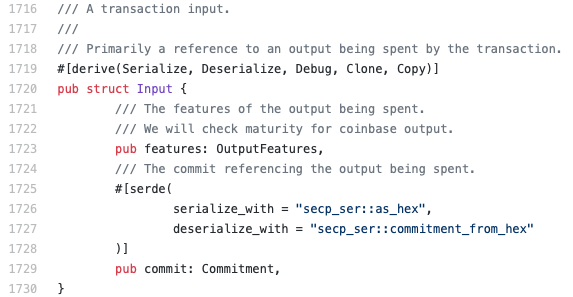}
	\caption{\texttt{Grin} input source code \cite{grin_github}}
	\label{fig:grin_input} 
\end{figure}

In order to provide privacy and confidentiality guarantees, \texttt{Grin} transactions are based on confidential transactions. In Figure \ref{fig:grin_transac}, we can observe that each transaction contains a list of inputs and outputs. Each input and output is in the form of a Pedersen commitment, i.e, a linear combination of the value of the transaction and a blinding factor. For instance, in the input structure (Figure \ref{fig:grin_input}, line 1729), there is a field that stores the commitment pointing to the output being spent.

In addition, the transaction structure has a list of transaction kernels (of type \textit{TxKernel}) with the transaction excess and the kernel signature. All this data has a straightforward relation to our definition of transaction (Definition \ref{def:transac}).

\begin{figure}
	\centering
	\includegraphics[width=0.65\textwidth]{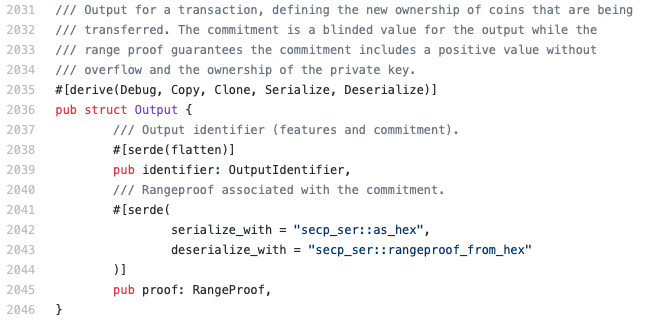}
	\caption{\texttt{Grin} output source code \cite{grin_github}}
	\label{fig:grin_output} 
\end{figure}

However, it is important to notice that the transaction kernel structure differs from our model since it does not contain the list of range proofs of the outputs. In \texttt{Grin}, it is part of the output structure (Figure \ref{fig:grin_output}, line 2045).

Moreover, a \texttt{Grin} transaction also includes the block number at which the transaction becomes valid. We did not add this data to the transaction structure yet. We also should include it in the signature process. In \texttt{Grin}, not only the transaction fee is signed, the signing process also takes into account the absolute position of the blocks in the chain. In this way, if a kernel block points to a height greater than the current one, it is rejected. If the relative position points to a specific kernel commitment, \texttt{Grin} has the same behavior.
 
\texttt{Grin} Blocks also stores a kernel offset which is the sum of all the transaction kernel offsets added to the block. In our model, the kernel offset is defined within a block (Definition \ref{def:block}) and the notion of adding a transaction into a block is formalized on the block aggregation (Definition \ref{def:aggregate_block}). Besides, the single aggregate offset allows to prove Lemma \ref{lemma_unlinkability} as part of the Transaction Unlinkability property (Property \ref{prop_unlinkability}).

%The kernel offset is used to hide which kernel belongs to which transaction and we only have a summed kernel offset stored in the header of each block.
%We don't have to record these transactions inside the block, although we still have to record the kernel as the kernel proof transfer of ownership to make sure that the whole block sums to zero, as expressed in the following formula:

\subsubsection{Privacy and Security Properties}

The cut-through process, as explained in Section \ref{sec:protocol_properties}, provides scalability and further anonimity. \texttt{Grin} performs this process in the transaction pool, which we formalized as \textit{mempool} (Definition \ref{def:mempool}).
Outputs which have already been spent as new inputs are removed from the mempool, using the fact that every transaction in a block should sum to zero.

%The following sourcecode shows how Grin implements this operation \cite{grin_github}:

CoinJoin, as we mentioned in Section \ref{sec:privacy_security_properties}, combines inputs and outputs from multiple transactions into a single transaction in order to obfuscate them. In \texttt{Grin}, every block is a CoinJoin of all other transactions in the block. 

In addition, \texttt{Grin} supports a \emph{pruning process}. This process could be applied to past blocks. Outputs that have been spent in a previous block are removed from the block. Block validity (Property \ref{prop:valid_block}) should be invariant w.r.t. the pruning process. Each node maintains a local state with a local copy of the chain. The pruning process can be applied recursively to the chain and keep it 
as compact as possible. Pruning is useful to free space. As a consequence, when a new 
node wants to join the network, it can receive just a pruned (i.e. partial) 
chain and the node needs to validate it, which makes the synchronization process faster. In Section \ref{secc:validating_chains}, $validChain$ should be modified to guarantee the validity of a partial chain.

As we have mentioned in Section \ref{sec:privacy_security_properties}, Switch commitments provides perfect hiddenness and statistical bindingness. \texttt{Grin} implements a switch commitment \cite{grin_switch_commitments} as part of a transaction output in order to provide more security than computational bindingness (Definition \ref{def:binding_commitment}), which is crucial for the age of quantum adversaries.

%Verification of zero sums. The sum of outputs minus inputs always equals zero, proving that the transaction did not create new funds, without revealing the actual amounts.

%Possession of private keys. Like with most other cryptocurrencies, ownership of transaction outputs is guaranteed by the possession of ECC private keys. However, the proof that an entity owns those private keys is not achieved by directly signing the transaction.

%Balance

%For example, Grin has implemented a method for a node to sync the blockchain very quickly by only downloading a partial history [11]. A new node entering the network will query the current head block of the chain and then request the block header at a horizon. In the example, the horizon is initially set at 5,000 blocks before the current head. The node then checks if there is enough data to confirm consensus. If there isn't consensus, the node will increase its horizon until consensus is reached. At that point, it will download the full Unspent Transaction Output (UTXO) set of the horizon block. This approach does introduce a few security risks, but mitigations are provided and the result is that a node can sync to the network with an order of magnitude less data.

%Grin unique features:
%Partial history syncing.
%DAG representation of Mempool to prevent duplicate UTXOs and cyclic transaction references.

%-- Schnorr signature --

%https://tlu.tarilabs.com/protocols/grin-protocol-overview/MainReport.html#appendix-a-example-of-grin-block

\subsection{Beam}
%https://docs.beam.mw/BEAM_Comparison_with_classical_MW.pdf

\texttt{Beam} \cite{beam_white_paper} was the other Mimblewimble project launched on January 2019. This open source system has a founding model and a dedicated development team.

%Athough it has a founding model and a dedicated development team, the code is still open source.
%On the other hand, Beam's project contains some modifications to the original approach providing some unique features. 

\subsubsection{Blocks and Transactions}

\texttt{Beam} transactions are confidential transactions implemented by the Pedersen commitment scheme. This follows the same approach as our model.

\begin{figure}
	\centering
	\includegraphics[width=0.53\textwidth]{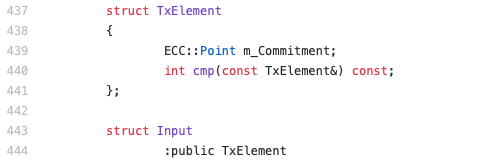}
	\caption{\texttt{Beam} input source code \cite{beam_github}}
	\label{fig:beam_input} 
\end{figure}

Figure \ref{fig:beam_input} shows (line 439) how \texttt{Beam}'s input stores the commitment, i.e, a point over the elliptic curve (class of \textit{ECC::Point}).

\begin{figure}
	\centering
	\includegraphics[width=0.7\textwidth]{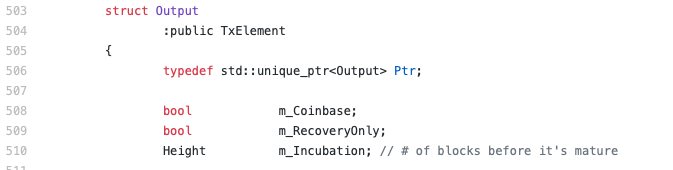}
	\caption{Output incubation period source code \cite{beam_github}}
	\label{fig:beam_incubation} 
\end{figure}

In Section \ref{sec:model} we described how each node maintains a local state. The state keeps track of the unspent transaction output set (UTXOs). \texttt{Beam} extends the behaviour of that set, supporting the incubation period on a UTXO. This means that \texttt{Beam} sets the minimum number of blocks created after the UTXO entered the blockchain, before it can be spent in a transaction. This number is included in the transaction signature. 
Figure \ref{fig:beam_incubation} shows (line 510) how \texttt{Beam}'s output stores the number of blocks corresponding to the incubation period.

In our model (Section \ref{secc:validating_chains}), the predicate $\mathit{validChain}$ should check that every output with certain incubation period on a block was `lawfully' spent for the entire blockchain (global state). In other words, if we have an output transaction $o$ with an incubation period $d$ on a confirmed block $b$ over the chain and a later confirmed block $b'$ containing $o$ as an input, then $b'$ should be, at least, $d$ blocks away from $b$ on the blockchain.

\subsubsection{Privacy and Security Properties}

\texttt{Beam} supports cut-through as we described above. In addition, \texttt{Beam} adds a scalable feature to eliminate all intermediate transaction kernels \cite{beam_mw} in order to keep the blockchain as compact as possible. 
It would be important to prove that the resulting transaction is still valid in Property \ref{prop:valid_transac}.

\subsection{Discussion}
Both \texttt{Grin} and \texttt{Beam} implementations address the main features of the MW protocol, namely the properties of  confidentiality, anonymity and unlinkability comprised in our work. 

\subsubsection{Broadcasting Protocol} Both \texttt{Grin} and \texttt{Beam} use the Dandelion scheme as broadcasting protocol \cite{DBLP:journals/corr/Venkatakrishnan17}. We have formalized that a broadcasting protocol should hold Property \ref{prop_untraceability} of Transaction Untraceability. 
It should not be possible to link transactions and their originating IP addresses, in other words, to deanonymize users. Broadcast with confusion, as we describe in Property \ref{prop:broadcast_conf}, should be carried out to satisfy Transaction Untraceability. We have also described the steam and fluff phases of the Dandelion scheme.

\texttt{Grin}'s implementation, in the steam phase, allows for transaction aggregation (CoinJoin) and cut-through, which provides greater anonymity to the transactions before they are broadcasted to the entire network.

In addition, in order to improve privacy, \texttt{Beam}'s implementation adds dummy transaction outputs at the steam phase. Each output has a value of zero and it is indistinguishable from regular outputs. Later, after a random number of blocks, the UTXOs are added as inputs to new transactions, i.e., they are spent and removed from the blockchanin.

In Section \ref{secc:unlinkability} we have specified the steam routine. Following, we extend the routine to capture \texttt{Beam}'s behaviour:

\vspace{2ex}

$\begin{array}{l}
subroutine\ steam(tx:Transaction) \{\\
\text{(* incubation period random choice *)} \\
i \leftarrow \{min, max\} \\
\text{(* create zero value output with incubation period i *)} \\
zeroOut \leftarrow createZeroValueOutput(i)  \\
addOutputTransactionUTXO(zeroOut) \\
addOutputTransaction(zeroOut, tx) \\
... \\
\}\\ \\
\end{array}$

To capture that semantic, we have combined two \texttt{Beam}'s features: incubation period on UTXO and aggregation of zero value transaction outputs. Firstly, we randomly choose $i$ as an incubation period. Then, we create a zero value transaction output ($zeroOut$) with incubation period $i$. The incubation period will ensure not to spend the dummy output before $i$ blocks are confirmed on the chain. After that, $zeroOut$ is added to the UTXO set (which is maintained in the local state of the node) and to the transaction $tx$ that is being broadcasted. Finally, the routine follows as we specified in Section \ref{secc:unlinkability}.

\subsubsection{Range proofs} 

\texttt{Grin} and \texttt{Beam} implement range proofs using Bulletproofs \cite{BunzBullet}. Bulletproofs are a non-interactive zero-knowledge proof protocol. They are short proofs (logarithmic in the witness size) with the aim of proving that the committed value is in a certain (positive) range without reveling it. Proof generation and verification times are linear in the length of the range. Regarding our model, it is the first property a transaction should satisfy to be valid (Property \ref{prop:valid_transac}). Furthermore, for every transaction in a bock, the range proofs of all the outputs should be valid too (Property \ref{prop:valid_block}).

\subsubsection{Some Design Decisions}

\paragraph{\textit{Emission Scheme}} It is known that BitCoin has a limited and finite supply of coins. Nowadays, new coins come from the process called ``mining'' where miners are paid because of their work of aggregating new blocks to the chain besides of the transaction fees. However, once the maximum amount of coins in circulation is reached, there will not be new coins and the miners will be paid only with the transaction fees.

\texttt{Grin} has a different approach. It has a static emission rate, where a fixed number of coins is released as a reward for agreggating a new block to the chain. This algorithm has no upper bound for new coins. 
However, \texttt{Beam} has a capped total supply standing at 262M. The reward algorithm is decreasing over the years \cite{beam_emission}.

\paragraph{\textit{Parties Negotiation}} Mimblewimble establishes that communication between the parties to construct a new transaction is made off-chain. Parties should collaborate in order to choose blinding factors and construct a valid transaction, in particular, a balanced transaction as in Definition \ref{def:balanced_transac}. 
\texttt{Grin} offers this process synchronously. Both parties are connected directly to one another and they should be online simultaneously.

On the other hand, in order to construct a new transaction, \texttt{Beam} offers a non-interactive negotiation between the parties. The Secure Bulletin Board System (SBBS) \cite{sbbs} runs on the nodes and it allows the parties to communicate off-line. Moreover, \texttt{Beam} also presents a one-side payment scheme. This scheme allows senders to pay a specified value to a particular receiver, without any interaction from the receiver side. The key here is not revealing blinding factors. It is addressed with a process called kernel fusion. Basically, both parties construct a half kernel and both kernels should be present in the transaction.

\paragraph{\textit{Chain Syncronization}} \texttt{Grin} allows partial history syncronization. When a new node wants to enter the network, it is not necesary to download the full history of the chain but it will query the block header at a horizon. The node can increase this limit as necessary. Then, it will download the full UTXO set of the horizon block.

\texttt{Beam} improves node synchronization using macroblocks. A macroblock is a compressed version of blockchain history after applying the cut-through process. Each node stores macroblocks locally. When a new node connects to the network, it will download the latest macroblock and will start working from that point. 

%https://documentation.beam.mw/en/latest/rtd_pages/user_glossary.html

%\paragraph{Proof of Work}  

%The BEAM team opted for modified Equihash PoW (Equihash 150,5 with additional datapath change intended to deter ASICs). This, in addition to a planned six-month hard fork, is intended to deter ASICs from the network.

%Grin uses the novel Cuckoo Cycle (to start with), an alternative proof of work system developed by John Tromp in 2015. While the algorithm is intended to be ASIC resistant, the team has long believed ASICs are a potentially inevitability. The solution emerged in September 2018: using two algorithms, one optimized to be ASIC-friendly (Cuckatoo31+) and another to allow GPUs to compete (Cuckaroo29), where the PoW balances mining rewards between the two every 24-hour period.

%Ownership

%In the previous section we introduced a private key as a blinding factor to obscure the transaction's values. The second insight of Mimblewimble is that this private key can be leveraged to prove ownership of the value.

%Alice sends you 3 coins and to obscure that amount, you chose 28 as your blinding factor (note that in practice the blinding factor, being a private key, is an extremely large number). Somewhere on the blockchain, the following output appears and should only be spendable by you:

%\input{codeverif}
%\input{prototype}
\section{Final remarks}
\label{sec:conclusion}
MW constitutes an important step forward in the protection of anonymity and privacy in the domain of cryptocurrencies.
Since it facilitates traceability and the validation process, both \texttt{Grin} and \texttt{Beam} adopted the MW protocol for their implementations. 
%Both cryptocurrencies do not feature addresses, transaction amounts and a record of all transaction histories. \hcomm{por que es importante remarcar esta carencia?}

We have highlighted elements that constitute essential steps towards the development of an exhaustive formalization  of the MW cryptocurrency protocol, the analysis of its properties  and the verification of its implementations. The proposed idealized model is key in the described verification process. We have also identified and precisely stated sufficient conditions for our model to ensure the verification of relevant security properties of MW. 
With respect to our previous paper~\cite{BetarteCLSZ20}, we have extended the definition of the MW protocol and the idealized model, incorporating in particular the discussion on the security properties of Pedersen commitments. Furthermore, we have studied the strength of the commitment scheme introduced regarding the main security properties a cryptocurrency protocol must have.

Since MW is built on top of a consensus protocol, we have developed a Z specification of a consensus protocol and presented an excerpt of the \setlog prototype after its Z specification.  This \setlog prototype can be used as an executable model where simulations can be run. This allows us to analyze the behavior of the protocol without having to implement it in a low level programming language.

Finally, we analyze and compare the \texttt{Grin} and \texttt{Beam} implementations in their current state of development, considering our model and its properties as a reference base.

We plan to continue working on the lines presented in Section \ref{sec:verification-MW}, namely,  considering tools oriented towards the verification of cryptographic protocols and implementations, such as \texttt{EasyCrypt} \cite{DBLP:conf/fosad/BartheDGKSS13}, \texttt{ProVerif} \cite{DBLP:conf/csfw/Blanchet01}, \texttt{CryptoVerif} \cite{BlanchetDagstuhl07} and \texttt{Tamarin} \cite{tamarin}. In particular, we are especially interested in using \texttt{EasyCrypt}\footnote{See \url{http://www.easycrypt.info}.}, an interactive framework for verifying the security of cryptographic constructions in the computational model.

%We plan to develop a complete and uniform formulation of several security properties of the protocol using the \texttt{Coq} proof assistant \cite{coq-manual}. \texttt{Coq} has an important set of libraries; for example \cite{DBLP:conf/itp/BartziaS14} contains a formalization of elliptic curves theory, which allows the verification of elliptic curve cryptographic algorithms.
%Applying the program extraction mechanism provided by Coq we have also propose to derive a certified Haskell prototype of the protocol, which can be used as a testing oracle and also to conduct verification activities on actual implementations of the protocol.

\bibliography{biblio}  
%\newpage
\small
\appendix

\section{\label{backz}Background on the Z Notation}
The Z notation is a formal method based on first-order logic and Zermelo-Fraenkel set theory. Any Z specification takes the form of a state machine---not necessarily a finite one. This machine is defined by giving its state space and the transitions between those states. The state space is given by declaring a tuple of typed state variables. A transition, called operation in Z, is defined by specifying its signature, its preconditions and its postconditions. The signature of an operation includes input, state and output variables. Each operation can change the state of the machine. State change is described by giving the relation between before-state and after-state variables. 

In order to introduce the Z notation we will use a simple example. Think in the savings accounts of a bank. Each account is identified by a so-called account number. The bank requires to keep record of just the balance of each account. Since account numbers are used just as identifiers we can abstract them away, not caring about their internal structure. Z provides so-called basic or given types for these cases. The Z syntax for introducing a basic type is:
\begin{zed}
[ACCNUM]
\end{zed}
In this way, it is possible to declare variables of type $ACCNUM$ and it is possible to build more complex types involving it---for instance the type of all sets of account numbers is $\power ACCNUM$.

We represent the money that clients can deposit and withdraw and the balance of savings accounts as natural numbers. The type for the integer numbers, $\num$, is built-in in Z. The notation also includes the set of natural numbers, $\nat$. Then, we define:
\begin{zed}
MONEY == \nat \also
BALANCE == \nat
\end{zed}
In other words, we introduce two synonymous for the set of natural numbers so the specification is more readable.

The state space is defined as follows:
\begin{schema}{Bank}
balances: ACCNUM \pfun BALANCE \\
\end{schema}
This construction is called schema; each schema has a name that can be used in other schemas. In particular this is a state schema because it only declares state variables. In effect, it declares one state variable by giving its name and type. Each state of the system corresponds to a particular valuation of this variable. The type constructor $\pfun$ defines partial functions\footnote{It must be noted that the Z type system is not as strong as the type systems of other formalisms, such as Coq. So we will be as formal as is usual in the Z community regarding its type system.}. Then, $balances$ is a partial function from $ACCNUM$ onto $BALANCE$. It makes sense to define such a function because each account has a unique $ACCNUM$; and it makes sense to make $balances$ partial because not every account number is used all the time at the bank (i.e. when an account is opened the bank assigns it an unused account number; and when an account is closed its account number is no longer used).

Now, we can define the initial state of the system as follows:
\begin{schema}{InitBank}
Bank
\where
balances = \emptyset
\end{schema}
$InitBank$ is another schema. The upper part is the declaration part and the lower part is the predicate part---this one is optional and is absent from $Bank$. In the declaration part we can declare variables or use schema inclusion. The latter means that we can write the name of another schema instead of declaring their variables. This allows us to reuse schemas. In this case the predicate part says that the state variable is equal to the empty set. It is important to remark that the $=$ symbol is logical equality and not an imperative assignment---Z has no notion of control flow. In Z, functions are sets of ordered pairs. Being sets they can be compared with the empty set. The symbol $\emptyset$ is polymorphic in Z: it is the same for all types. 

Now we can define an operation of the system (i.e. a state transition) specifying how an account is opened.
\begin{schema}{NewAccountOk}
\Delta Bank \\
n?:ACCNUM \\
\where
n? \notin \dom balances \\
balances' = balances \cup \{n? \mapsto 0\}
\end{schema}
As can be seen, operations are defined by means of schemas. The expression $\Delta Bank$ in the declaration part is a shorthand for including the schemas $Bank$ and $Bank'$. We already know what it means including $Bank$. $Bank'$ is equal to $Bank$ but all of its variables are decorated with a prime. Therefore, $Bank'$ declares $balances'$ of the same types than that in $Bank$. When a state variable is decorated with the prime it is assumed to be an after-state variable. The net effect of including $\Delta Bank$ is, then, the declaration of one before-state variable and one after-state variable. A $\Delta$ expression is included in every operation schema that produces a state change.

Variables decorated with a question mark, like $n?$, are assumed to be input variables. Then, $n?$ represents the account number to be assigned to the new savings account. To simplify the specification a little bit we assume that a bank's clerk provides the account number when the operation is called---instead of the system generating it.

Note that the predicate part consists of two atomic predicates. When two or more predicates are in different rows they are assumed to be a conjunction. In other words:
\[
n? \notin \dom balances \\
balances' = balances \cup \{n? \mapsto 0\}
\]
is equivalent to:
\[
n? \notin \dom balances \land
balances' = balances \cup \{n? \mapsto 0\}
\]

Z uses the standard symbols of discrete mathematics and set theory so we think it will not be difficult for the reader to understand each predicate. Remember that functions and relations are sets of ordered pairs so they can participate in set expressions. For instance, $balances \cup \{n? \mapsto 0\}$ adds an ordered pair to $clients$. Again, the expression $balances' = balances \cup \{n? \mapsto 0\}$ is actually a predicate saying that $balances'$ is equal to $balances \cup \{n? \mapsto 0\}$, and not that the latter is assigned to the former. In other words, this predicate says that the value of $balances$ in the after-state is equal to the value of $balances$ in the before-state plus the ordered pair $n? \mapsto 0$.

Note that operations are defined by giving their preconditions and postconditions. In $NewClientOk$ the precondition is:
\[
n? \notin \dom balances
\]
\noindent while its postcondition is:
\[
balances' = balances \cup \{n? \mapsto 0\}
\]

Therefore, $NewClientOK$ does not say what the system shall do when $n? \notin \dom balances$ does not hold. The bank says that nothing has to be done when the account number chosen by the clerk is already in use. Then, we define a new schema for this case:
\begin{zed}
AccountAlreadyExists == \\
   \t1 [\Xi Bank; n?:ACCNUM | n? \in \dom balances]
\end{zed}
This is another way of writing schemas, called horizontal form. It has the same meaning than:
\begin{schema}{AccountAlreadyExists}
\Xi Bank \\
n?:ACCNUM 
\where
n? \in \dom balances
\end{schema}

The expression $\Xi Bank$ is a shorthand for:
\begin{schema}{\Xi Bank}
\Delta Bank
\where
balances' = balances
\end{schema}

If a $\Xi$ expression is included in an operation schema, it means that the operation will not produce a state change because all the primed state variables are equal to their unprimed counterparts. When a schema whose predicate part is not empty is included in another schema, the net effect is twofold: (a) the declaration part of the former is included in the declaration part of the latter; and (b) the predicate of the former is conjoined to the predicate of the latter. Hence, $AccountAlreadyExists$ could have been written as follows:
\begin{schema}{AccountAlreadyExists}
balances, balances': ACCNUM \pfun BALANCE \\
n?:ACCNUM 
\where
n? \in \dom balances
balances' = balances
\end{schema}

Usually, schemas like $NewClientOk$ are said to specify the successful cases or situations, while schemas like $AccountAlreadyExists$ specify the erroneous cases. Finally, we assemble the two schemas to define the total operation---i.e. an operation whose precondition is equivalent to $true$---for opening a savings account in the bank:
\begin{zed}
NewClient == NewClientOk \lor AccountAlreadyExists
\end{zed}

$NewClient$ is defined by a so-called schema expression. Schema expressions are expressions involving schema names and logical connectives. Let $A$ be the schema defined as $[D_A | P_A]$ where $D_A$ is the declaration part and $P_A$ is its predicate. Similarly, let $B$ the schema defined by $[D_B | P_B]$. Then, the schema $C$ defined by $A \circledast B$, where $\circledast$ is any of $\land$, $\lor$ and $\implies$, is the schema $[D_A; D_B | P_A \circledast P_B]$. In other words, the declaration parts of the schemas involved in a schema expression are joined together and the predicates are connected with the same connectors used in the expression---if there is some clash in the declaration parts it must be resolved by the user. In symbols:
\[
A == [D_A | P_A] \also
B == [D_B | P_B] \also
C == A \circledast B \text{, where } \circledast \text{ is any of } \land, \lor, \implies \text{ then} \also
C == [D_A; D_B | P_A \circledast P_B]
\]

\end{document}